\documentclass[numberedappendix,apj]{emulateapj}
\usepackage{amssymb}
\usepackage{amsmath}

\shorttitle{Turbulent Mixing Layers}
\shortauthors{Esquivel et al.}
\slugcomment{Draft version, \today}

\begin{document}

\title{MHD Turbulent Mixing Layers:  Equilibrium Cooling Models}

\author{
Alejandro Esquivel\altaffilmark{1,2},
Robert A. Benjamin\altaffilmark{3},
Alex Lazarian\altaffilmark{1},\\
Jungyeon Cho\altaffilmark{4},
and Samuel N. Leitner\altaffilmark{5}
}
\altaffiltext{1}{Astronomy Department, University of
  Wisconsin-Madison, 475 N.Charter St., Madison, WI 53706, USA}
\altaffiltext{2}{Instituto de Ciencias Nucleares, Universidad Nacional
  Aut\'{o}noma de M\'{e}xico, Apartado Postal 70-503, 04510 M\'{e}xico
  D.F., M\'{e}xico}
\altaffiltext{3}{Department of Physics, University of
  Wisconsin-Whitewater, Whitewater, WI 53190, USA}
\altaffiltext{4}{Department of Astronomy and Space Science, Chungnam
  National University, 220 Kung-dong, Yusong-ku, Daejon 305-764,
  Korea}
\altaffiltext{5}{Department of Physics, Wesleyan University,
  Middletown, CT 06459, USA}

\email{esquivel@astro.wisc.edu}

\begin{abstract}

We present models of turbulent mixing at the boundaries between hot
($T\sim 10^{6-7}~\rm{K}$) and warm material ($T\sim 10^4~\rm{K}$) in the
interstellar medium, using a three-dimensional magnetohydrodynamical
code, with radiative cooling. The source of turbulence in our
simulations is a Kelvin-Helmholtz instability, produced by shear
between the two media. We found, that because the growth rate
of the large scale modes in the instability is rather slow, it takes a
significant amount of time ($\sim 1~\rm{Myr}$) for turbulence to
produce effective mixing.  We find that the total column densities of
the highly ionized species (\ion{C}{4}, \ion{N}{5}, and \ion{O}{6})
per interface (assuming ionization equilibrium)  are similar to
previous steady-state non-equilibrium ionization models, but grow
slowly from 
$\log N \sim 10^{11}$ to a few $ \times 10^{12}~{\rm cm^{-2}}$ as the
interface evolves. However, the column density ratios can differ
significantly from previous estimates, with an order of magnitude
variation in  N(\ion{C}{4})/N(\ion{O}{6}) as the mixing develops.   

\end{abstract}


\section{Introduction}

The interstellar medium (ISM) is a very complex entity. It is extremely
rich in structure, highly turbulent, and embedded in a dynamically
important magnetic field.
Although the concept of ``phases'' is up for debate \citep{C05}, we
can safely say that the ISM shows regions of distinct physical
conditions, which range from cold and molecular, to hot and ionized.
A factor that controls the structure of the ISM to a large extent, is
the balance between heating and cooling processes.
In this work, we are mainly concerned with the interfaces between the
{\it hot} ($T\sim 10^6$K) and {\it warm} ($T\sim 10^4$K) media, as
well as the transfer of heat between them.
At temperatures in between the hot and warm media the efficiency of
radiative cooling is maximal \citep[see][]{SD93}.
Material at such temperatures cools very quickly, therefore
should be rarely observed. However, absorption line studies in the far
ultraviolet (FUV) have found otherwise \citep[e. g.][]{S03},
suggesting substantial amounts of plasma at temperatures of $T\sim
10^5$K.

It was realized by \citet[][hereafter BF90]{BF90}, that the interfaces
between hot and warm media might well be dominated by ``turbulent
mixing layers''. The general idea was that turbulence produces an
exchange of material and energy at the boundary between the two media,
providing a steady supply of plasma at $T\sim 10^5$K, balancing
the losses due to radiation.
Later, \citet*[][hereafter SSB93]{SSB93} predicted that line ratios of
highly ionized species (\ion{C}{4}, \ion{N}{5}, \ion{O}{6}, and
\ion{Si}{4}) differ significantly from those in purely photoionized
gas, radiative shocks, and conduction fronts, thus providing a useful
diagnostic tool of the physical conditions in turbulent mixing
layers. Their model however, relied on several simplifying
assumptions. For instance they characterized the mixing layer with a
single mean temperature $\bar{T}$, regardless of the position in  the
mixing layer. The model is one-dimensional, unmagnetized, and
parameters were chosen to correspond observations. 
The obvious requirement for this type of heat transport is the
development of turbulence at the boundary of the two media. In the ISM
there are many outflows and plasma instabilities that can lead to
this situation. For example, in a supernova explosion Rayleigh-Taylor
instabilities occur when the hot and low-density ejecta tries to
accelerate a colder and denser medium. This produces a boundary layer
that breaks up into small (colder) high density clumps left inside a
hot cavity.
Another physical process that can lead to a turbulent mixing
layer is the Kelvin-Helmholtz (K-H) instability, which can occur when
shear is present between two fluids. This is possible, for instance, at
the edges of high velocity clouds (HVCs) moving through the galactic
hot corona \citep[see][]{WW97}.

However, the presence of a magnetic field can influence the dynamics
of the K-H instability. It is well known that flows with an Alfv\'{e}n
speed less than the shear velocity can become stable \citep[see for
  instance][]{C61}. A detailed numerical study of the K-H instability
can be found in \citet*{FJRG96,RJF00}. Enhancement of thermal
difussivity in magnetized plasmas due to turbulent motions is
discussed in \citet{CLHKKM03}.

In this work, we study the formation and evolution of turbulent mixing
layers by means of the K-H instability. We use a magnetohydrodynamic
(MHD) code, which includes radiative cooling. The layout of the paper
is as follows: in \S2 we provide with a brief review of the theory
behind turbulent mixing layers, in \S3 we describe the code and our
numerical setup. The results, including estimates of column densities
and line ratios of highly ionized species can be found in \S4,
followed by a summary in \S5.

\section{Turbulent mixing layers}

The idea of a turbulent mixing layer was proposed in BF90 as an
important heat transfer mechanism between two media at different
temperatures. The basic picture proposed is that turbulence at the 
boundary of the two fluids will provide a constant input to the
intermediate temperature mixture. If energy is conserved, this mixing
layer would grow  indefinitely, and eventually all of the material
will be at such intermediate temperature.
To prevent this they proposed a steady state in which
the energy lost by radiation (radiative cooling is most efficient
precisely at such intermediate temperatures) is balanced by a turbulent
heat flux into the mixing layer. As it is usual, most of the energy in
the turbulence is on the largest scales, and cascaded down
to a dissipative level. Thus the model of BF90 is basically a
three-phase steady  state fluid in which the losses due to radiation
are balanced with energy entrained by turbulence into the
intermediate temperature zone. In their model, the temperature of the
mixing layer is determined by mass flux balance:
\begin{equation}
\bar{T}=\frac{\dot{m}_{hot}T_{hot}+\dot{m}_{cold}T_{cold}}{\dot{m}_{hot}+\dot{m}_{cold}},
\label{eq:Tbar}
\end{equation}
where,  $\dot{m}_{hot}$ and $\dot{m}_{cold}$ are the mass flux rates
from the hot and from the cold phase into the layer, $T_{hot}$ and
$T_{cold}$ are the temperatures of the hot and cold phases
respectively. Since neither the heat transfer nor the mixing will
be perfect, BF90 introduced two efficiency factors: $\eta_{hot}$, the
fraction of mass and energy that is deposited by the hot medium to the
mixing layer, and $\eta_{cold}$, an efficiency for the hydrodynamic
mixing. With these factors the mass flux rates become
\begin{eqnarray}
\dot{m}_{hot} & = & \eta_{hot}\rho_{hot}v_t,
\label{eq:mdot_hot}\\
\dot{m}_{cold} & = & \eta_{cold}\left(\rho_{hot}\rho_{cold}\right)^{1/2}v_t, 
\label{eq:mdot_cold}
\end{eqnarray}
and the resulting intermediate temperature will be:
\begin{eqnarray}
\bar{T} & \approx&
\left[
  \frac{\eta_{hot}+\eta_{cold}\left(T_{cold}/T_{hot}\right)^{1/2}}
       {\eta_{cold}+\eta_{hot}\left(T_{cold}/T_{hot}\right)^{1/2}}
\right]
\left(T_{cold}\,T_{hot}\right)^{1/2}\nonumber\\
& \equiv & \xi \left(T_{cold}\,T_{hot}\right)^{1/2}.
\label{eq:Tbar2}
\end{eqnarray}
The definition of the two different efficiencies reveals what the
authors had in mind as the mechanism that provides the mixing. The
efficiency associated with the entrainment of hot gas simply
corresponds to an enthalpy flux $\case{5}{2}\eta_h p v_t$, where $p$
is the pressure and $v_t$ the turbulent velocity. This yields
eq.(\ref{eq:mdot_hot}), where $\rho_h$ is the mass density of the hot
medium and the contribution of turbulent kinetic energy has been
neglected. The cold gas is then pulled into the hot medium by
turbulent eddies at a rate  $\eta_c \rho_c l_c/t(l_c)$. Turbulent
eddies are formed on scales $l_c < l_t$, where $l_t$ is the total
thickness of the layer, and $t(l_c)$ can be thought as an eddy turnover
time for eddies of size $l_c$. BF90 used the Kelvin-Helmholtz growth
rate $t_{KH}(l_c)\sim (\rho_c/\rho_h)^{1/2} l_c/v_t$, where $\rho_c$ is
the mass density of the cold medium, leading to
eq.(\ref{eq:mdot_cold}). However, the form of the timescale $\sim
(\rho_c/\rho_h)^{1/2} l_c/v_t$ is not exclusive for the K-H
instability but suitable for fully developed turbulence in general.
The main uncertainty in the BF90 model lies in these efficiency
factors.
It is also derived assuming fully developed turbulence (i.e. rapid
mixing), and therefore does not include the effects of cooling to the
dynamical development of turbulence.

Later, SSB93 expanded on the ideas of BF90 and ran a grid of models
based on one-dimensional,  instantaneous, steady-state mixing,
characterized principally by $\bar{T}$, and $v_t$. These included the
effects of non-equilibrium ionization and self-photoionization of the
gas in the mixing layer, but adopted a somewhat ad hoc value for $\xi$
and the efficiencies ($\eta$'s).

In this work we focus on the dynamical formation and development of
the mixing layer. We do not include effects of  non-equilibrium or
self-photoionization, but we  measure the the actual $\bar{T}$, which
is result a continuous distribution of temperatures that range from
$T_{hot}$ to $T_{cold}$,\footnote{Instead of  $T_{cold}$ we   will
  call it $T_{warm}$ because our choice of parameters  are to coincide
  with the {\it warm} phase of the ISM (at $T\sim 10^4~\rm{K}$).}
instead of fixing any efficiency factor.

\section{Our model}

\subsection{The code}
We solve the following system of equations:
\begin{eqnarray}
\frac{\partial \rho    }{\partial t} + \nabla \cdot (\rho \mathbf{ v}) =0,
\label{eq:rho}  \\
\frac{\partial \mathbf{ v} }{\partial t} + \mathbf{ v}\cdot \nabla \mathbf{ v}
   +  \frac{1}{\rho}  \nabla p
   -\frac{ (\nabla \times \mathbf{ B})\times \mathbf{ B}}{4\pi \rho} =0,  \\
\frac{\partial p}{\partial t} +  \mathbf{ v}\cdot \nabla {p} +
   \gamma p \nabla \cdot \mathbf{ v} =0, \\
\frac{\partial \mathbf{ B}}{\partial t} -
     \nabla \times (\mathbf{ v} \times\mathbf{ B}) =0,   \label{eq:mag}
\end{eqnarray}
with $\nabla \cdot \mathbf{B}= 0$. Here $\rho$ is the mass density,
$\mathbf{v}$ is the velocity, $p$ is the pressure, and $\mathbf{ B}$
the magnetic field. We use an ideal gas equation of state
$p=(\gamma-1)\rho\,u$, where $\gamma$ is the ratio of the specific
heats ($\gamma = c_p/c_v$), and $u$ is the specific internal energy.
We solve equations (\ref{eq:rho})-(\ref{eq:mag}) using a MUSCL-type
scheme with HLL fluxes \citep*{HLL83} and monotonized central limiter
\citep*[see][]{KNP01}, on a regular Cartesian grid.
We use the the flux-interpolated constrained transport (or ``flux-CT")
scheme \citep{T00} to enforce the divergence free condition of the
magnetic field.
A similar version of the code was used by \citet*{GMKT03} for
relativistic MHD calculations.
Our code gives satisfactory results for standard shock tube tests
\citep[see, for example][]{BW88,RJ95}.
When compared with our earlier hybrid ENO code (see an isothermal
version in \citealt{CL02}),  our current code runs faster, and yet
gives consistent results.
The overall scheme is second-order accurate.
After updating the system of equations along $x1$ direction,
we repeat similar procedures for $x2$ and $x3$ directions, with the
appropriate rotation of indexes. We use a two-stage Runge-Kutta method
for time integration.  The cooling at a given temperature is assumed
to be of that of plasma in collisional ionization equilibrium, with
solar metallicity, as given  in \citet*{BBC01}.
We consider $\gamma =5/3$, and solve the time evolution of specific
internal energy. For the runs with cooling, this is applied before
updating equations (\ref{eq:rho})-(\ref{eq:mag}), using an implicit
scheme (i.e. we simply subtract the internal energy that will
be lost in the time-step $\delta t$).  We do not explicitly
include heating due to absorption of radiation or thermal conduction.
The time-step for the MHD part is determined by the standard ``Courant''
condition. When we include cooling, we monitor the minimal cooling
time in the computational box and compare it with that of the MHD
part. If the minimum cooling time is shorter, it replaces the
time-step. However, this was very rarely the case, and typically the
dynamical time was shorter than the cooling time (by one or two orders
of magnitude in most cases).

We do not include thermal conduction explicitly. The diffusion of
heat at small scales is determined in our simulations by the numerical
diffusion, whose properties may be different from those of diffusion in
the actual interstellar gas. However, this does not greatly affect our
final results if the heat transport is determined by turbulence.
A discussion about turbulent and thermal conduction in a magnetized
medium can be found in \citet{L06}.
Turbulence provides an effective diffusion coefficient $\sim 1/3 v_{turb}
L_{inj}$, where $v_{turb}$ is the turbulent velocity, and $L_{inj}$
is the scale of energy injection, corresponding to the size of the
largest eddies \citep{CLHKKM03}.
We will see that in our models the injection scale is
utterly determined by the size of our computational box.
If the turbulent diffusion coefficient is much larger than the thermal
diffusion coefficient the heat transfer is dominated by turbulence.
An important property of turbulent heat transfer, as well
as the other transport processes, is that they do not depend on the
microscopic diffusivity. Indeed, provided that the gas is turbulent,
the mixing and the heat transport are happening approximately over one
large eddy turnover time. If the diffusivity on the atomic level
decreases, the turbulent cascade goes to yet smaller scales ensuring
heat transport that still scales as above. For typical parameters of
turbulent mixing layers in our models (that is $T\sim 10^5~\rm{K}$,
$n\sim 10^{-2}~\rm{cm^{-3}}$, $L_{inj}\sim 10~\rm{pc}$, $V_{turb}\sim
20~\rm{km s^-1}$), one obtains a Spitzer thermal diffusion coefficient
of $\kappa_{Sp}\lesssim 10^{24}~\rm{cm^{2}}~s^{-1}$ and a turbulent
diffusion coefficient of $\kappa_{turb}\gtrsim
10^{25}~\rm{cm^{2}}~s^{-1}$.
The difference is modest, however the presence of a magnetic field
will further suppress thermal conduction.
Moreover, the magnetic field does not have to be dynamically
dominant to produce an important effect on the thermal conductivity. 
And, in some sense, our unmagnetized models are equivalent to models
with a very weak magnetic field (ubiquitous in the ISM), that is
sufficient to suppress thermal conductivity significantly.
At the beginning of the simulations turbulence is not fully developed
and the turbulent diffusivity is small, but at the same time the
magnetic field is aligned with the interface, dramatically decreasing
electron conduction as well.
At more evolved stages however, turbulence will develop and thermal
conduction will be less suppressed as the magnetic field becomes
entangled. Estimates by \citet{NM01} suggest that for fully developed
turbulence the thermal conductivity is decreased by a factor of $\sim
5$ from the Spitzer value. Their model is, however, rather restrictive
as only turbulence with Alfv\'{e}n Mach number ($\mathcal{M}_A$) equal
to unity is discussed. For both $\mathcal{M}_A$ much larger, and much
smaller than unity the electron thermal conductivity is less
\citep[see][]{L06}.

\subsection{The numerical setup}

\begin{figure}
\epsscale{1.0}
\plotone{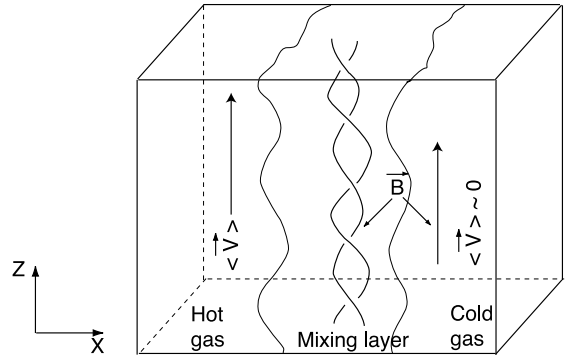}
\caption{Schematic of the calculation cube, with hot gas moving
  upward on the left, and initially static warm gas to the right.
\label{fig:sketch}}
\end{figure}

We start with warm gas, at a temperature of $T_{warm}=10^4~\rm{K}$,
with a relatively high hydrogen number density
$n_{warm}=0.1~\rm{cm^{-3}}$, and no mean motion, at the right side of
the computational box. At the opposite side we have
a low-density ($n_{hot}$), hot medium  ($T_{hot}$), moving upward ($z$
direction) with a velocity $v_{hot}$. In some of our simulations we
include a magnetic field $B_{z,warm}$ threading the warm gas aligned
in the vertical direction, with a magnitude corresponding to a
$\beta_{warm}\sim P_{gas}/P_{mag}\sim 1$. A sketch of the
computational domain is presented in Figure \ref{fig:sketch}.

The two media are initially in total pressure equilibrium. The
transition region (between hot and warm media) follows a hyperbolic
tangent profile in the $x$ direction, that initially occupies $\sim
1/10$th. of the box size. The division line for the runs in a $144^3$
grid was set at the middle of the box. The vertical direction in all
cases correspond to a physical scale of $L_{z}=10~\rm{pc}$. 
In the cases without cooling, the simulations can be rescaled
arbitrarily, but when cooling is included the scale lengths are fixed
by the value of the cooling rate used.
The boundary conditions are periodic in the $z$ and $y$
directions. For the $x$ direction in the warm side of the box (right)
we have a reflective boundary, and on the hot side (left) we
have a source boundary condition. The latter reinforces the
initial condition after each time-step, acting as a reservoir of hot
material, that helps to balance the energy lost trough radiation.
We initialized the computational cube with sinusoidal perturbations
($32$  harmonics, with random phases) in the component of the velocity
normal to the interface of the two media ($v_x$). The shear at the
boundary, will excite a K-H instability, which will eventually lead to
the development of turbulence.
 A summary of the parameters we use is presented in Table
\ref{tb:param}.
\begin{deluxetable*}{lccccc}
\tablecaption{ Parameters (initial conditions) of the runs used$^a$.
\label{tb:param}}
\tablewidth{0pt}
\tablehead{
\colhead{Model$^b$} & \colhead{$\log(T_{hot})$} 
& \colhead{$n_{hot}$}& \colhead{$v_{hot}$}
&\colhead{$B_{z,warm}$} & \colhead{Grid size} \\
\colhead{} & \colhead{$[\rm{K}]$} 
& \colhead{$[\rm{cm^{-3}}]$}& \colhead{$[\rm{km~s^{-1}}]$}
&\colhead{$[\mu\rm{G}]$} & \colhead{}
}
\startdata
144-Th6-V50-B0  & $6$ & $1\times10^{-3}$ & $50$  & $0$       & $144^3$ \\
144-Th6-V200-B0 & $6$ & $1\times10^{-3}$ & $200$ & $0$       & $144^3$ \\
256-Th6-V200-B0 & $6$ & $1\times10^{-3}$ & $200$ & $0$       & $256^3$ \\
LX-Th6-V200-B0  & $6$ & $1\times10^{-3}$ & $200$ & $0$       & $256\times144^2$ \\

144-Th7-V50-B0  & $7$ & $1\times10^{-4}$ & $50$  & $0$       & $144^3$ \\
144-Th7-V200-B0 & $7$ & $1\times10^{-4}$ & $20$  & $0$       & $144^3$ \\

144-Th6-V50-B1  & $6$ & $2\times10^{-3}$ & $50$  &  $\sim 2$ & $144^3$ \\
144-Th6-V200-B1 & $6$ & $2\times10^{-3}$ & $200$ &  $\sim 2$ & $144^3$ \\
144-Th7-V50-B1  & $7$ & $2\times10^{-4}$ & $50$  &  $\sim 2$ & $144^3$ \\
144-Th7-V200-B1 & $7$ & $2\times10^{-4}$ & $200$ &  $\sim 2$ & $144^3$ \\
\enddata
\tablenotetext{a}{ All the runs start with a temperature on the warm
side of $T_{warm}=10^4~\rm{K}$ and a hydrogen density of
$n_{warm}=0.1~\rm{cm^{-3}}$. The box size in the vertical direction
corresponds to $10~\rm{pc}$.}
\tablenotetext{b}{ All models were run either without cooling or with
  collisional ionization equilibrium cooling (see text for details).
  An additional prefix ``NC'' (no cooling) or ``EC'' (equilibrium
  cooling) to the run name is added respectively.}
\end{deluxetable*}

\section{Results}

\subsection{Evolution of the mixing layer}

\begin{figure}
\plotone{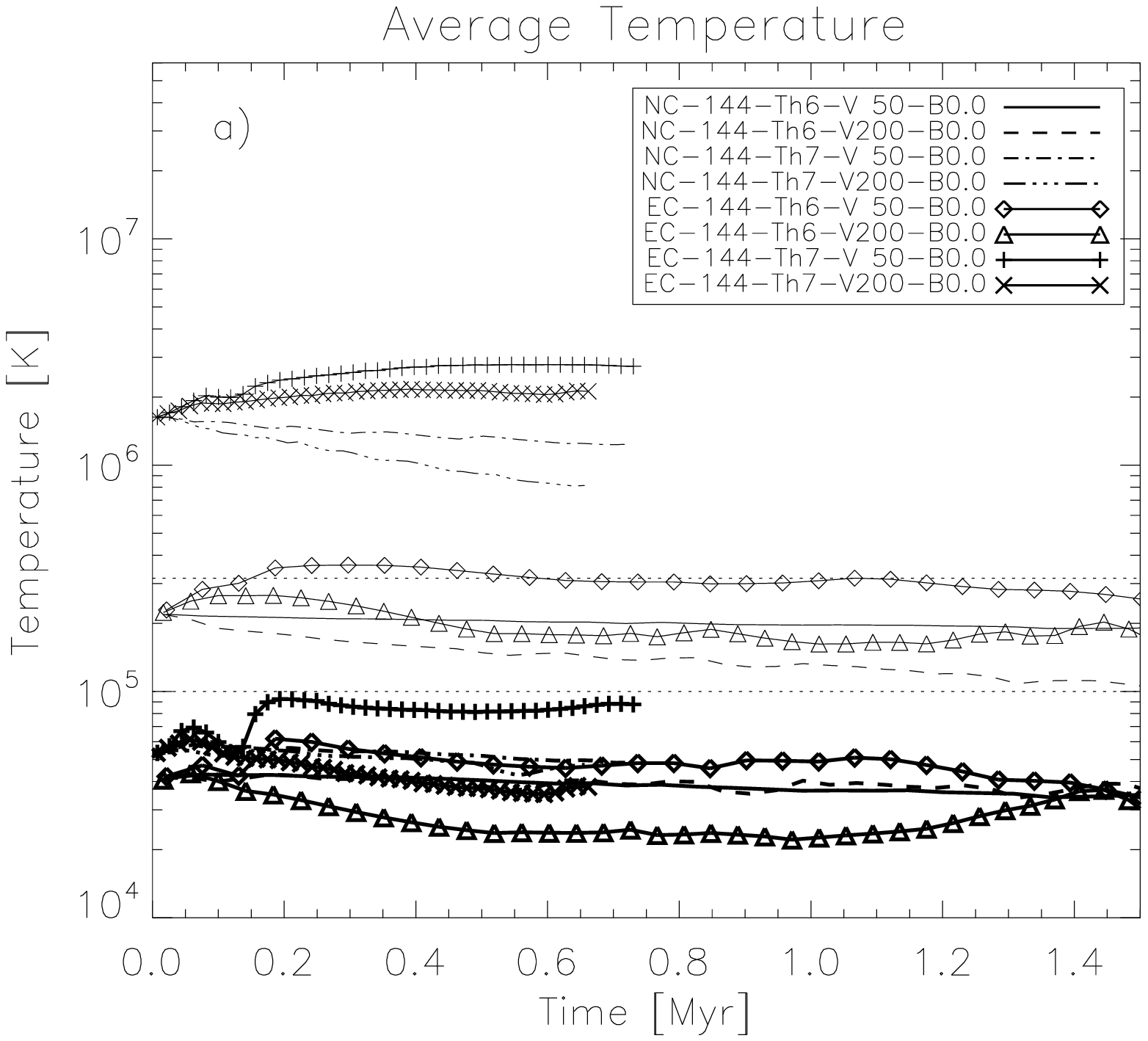}
\plotone{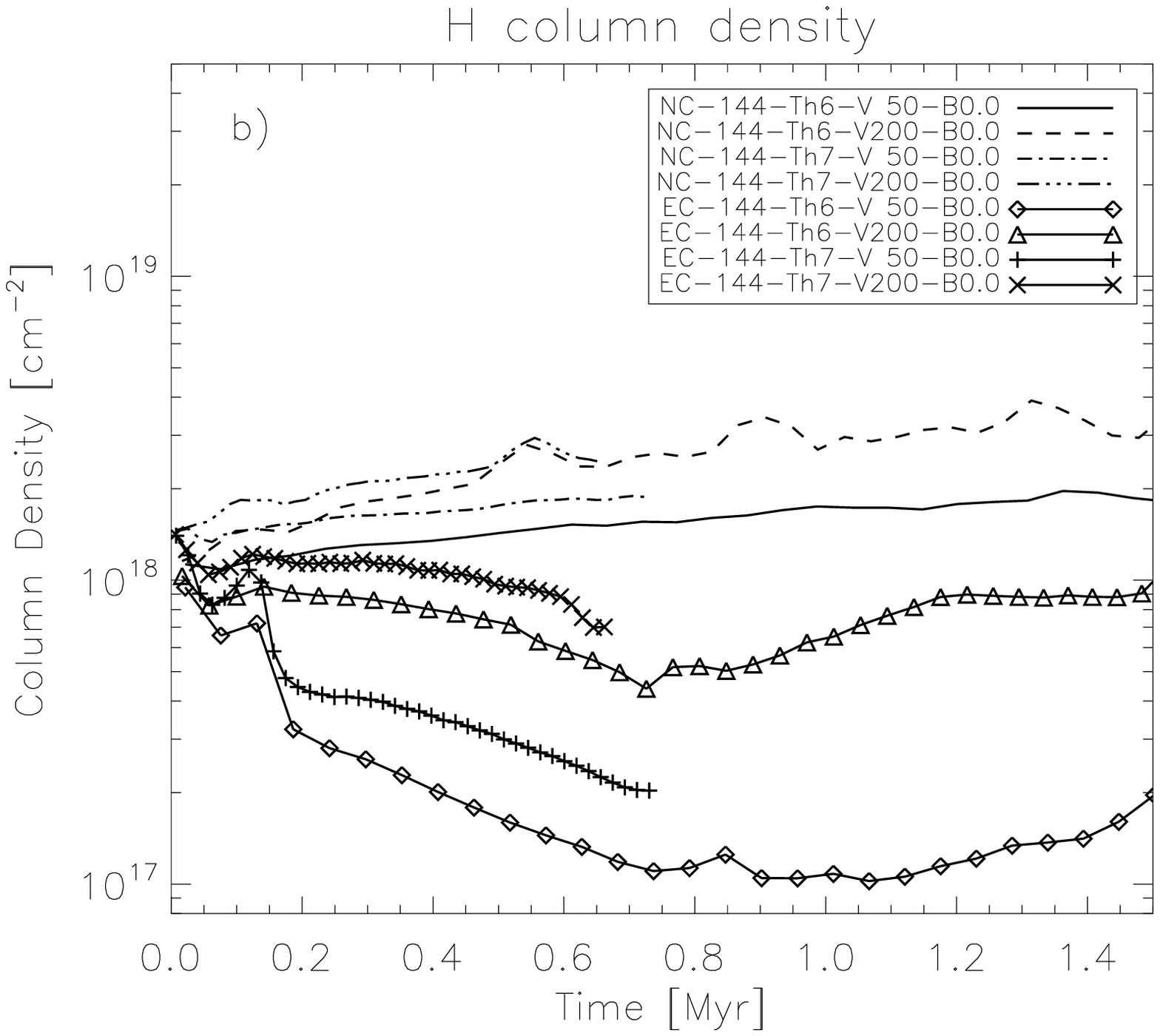}
\caption{Time evolution of the mixing layers, for the purely
  hydrodynamical runs. At the top (panel {\it a}) we present the
  density weighted average ({\it thick} lines), and  volume average
  ({\it thin} lines) temperature of the mixing layers (material with a
  departure of   $20\%$ from the nominal hot and warm
  temperatures). For reference we show the harmonic mean of the hot
  and warm temperatures (which corresponds to $\bar{T}$ in BF90, SSB93
  with $\xi=1$ ({\it dotted}) horizontal lines, see
  eq.[\ref{eq:Tbar}]). At the bottom (panel {\it b}) the mean hydrogen
  column density in the mixing layer (both panels are averages over
  different lines of sight normal to the layer  --along the $x$ axis--). 
\label{fig:t_ev_b0}}
\end{figure}

\begin{figure}
\plotone{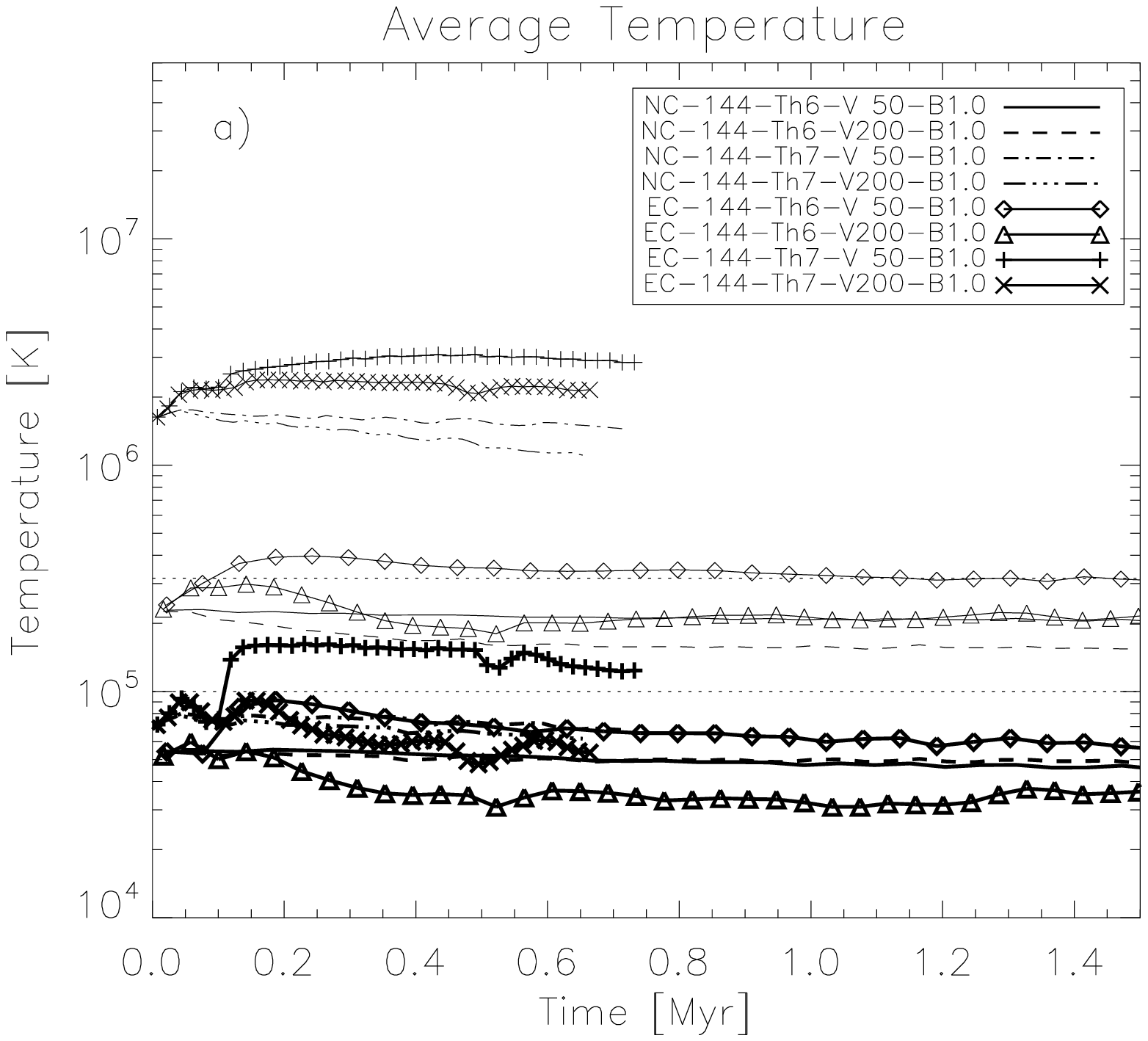}
\plotone{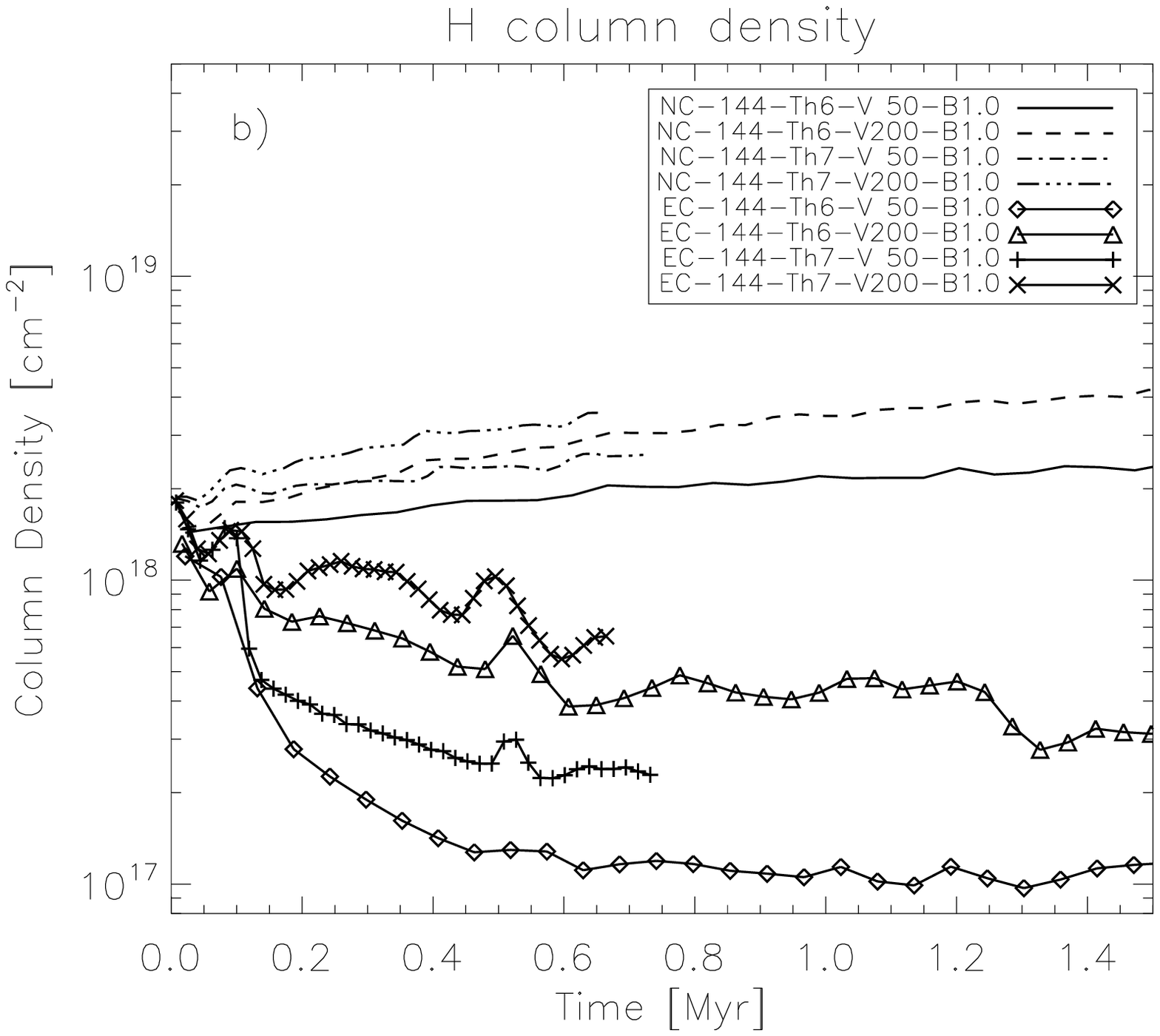}
\caption{ Same as Fig. \ref{fig:t_ev_b0}, for the cases with
  magnetic field ($\beta \sim 1$) on the warm side.
\label{fig:t_ev_b1}}
\end{figure}

Our models, as described in the previous section, are dynamical and
include a continuous transition between the {\it hot} and the
{\it warm} media.
In order to compare the result of our calculations with previous
models of turbulent mixing layers (SSB93) we have to define what a
region of ``intermediate temperature'' is, in our computational box.
We adopted a threshold of $20\%$ departure from any of the nominal
temperatures for the hot and warm media. That is, we consider in the
transition  region all material above $1.2\times 10^4~\rm{K}$, and
below $8\times 10^5~\rm{K}$ for the runs where $T_{hot}=1\times
10^6~\rm{K}$; or bellow $8\times10^6~\rm{K}$ for the runs with
$T_{hot}=1\times 10^7~\rm{K}$.
We evolve the initial conditions for $\sim$ 20,000 time-steps
for all the cases with $144^3$ resolution.
In Figures \ref{fig:t_ev_b0}, and \ref{fig:t_ev_b1} we show the time
evolution of all the purely hydrodynamical, and magnetized runs,
respectively.

We find in general, that the volume average temperature within the
mixing layer depends strongly on the threshold used to define the
intermediate temperature region.
The density weighted temperature is more robust against the particular
choice of threshold. And, since the emission  observable is also
density weighted, it is a better measure of the properties of mixing
layers, and we will use it to interpret our results.
Keeping in mind that the temperature in BF90 and SSB93 is a
simple volume average, and that denser regions correspond to lower
temperatures, our density weighted temperature will in general be
lower.
In our approach we do not assume a particular value for the
efficiencies ``$\eta$'s, and/or $\xi$'' , instead they are being
implicitly calculated. In principle they can be measured from the
simulations, however, it would require the choice of an arbitrary
temperature threshold to obtain a mean (volume average) temperature,
which turns out to be  very sensitive to such threshold. Thus, it is
impractical to compare the models through estimates of the
efficiencies

\subsubsection{Formation of the mixing layer: early stages}

Using a density weighted average, our calculations show a
relatively cold boundary layer (of a few $10^4~\rm{K}$) at the
early stages of formation. Due to the fact that the mass is initially
on the warm side, this indicates a somewhat inefficient mixing at
such early times.

\begin{figure*}
\epsscale{0.9}
\plottwo{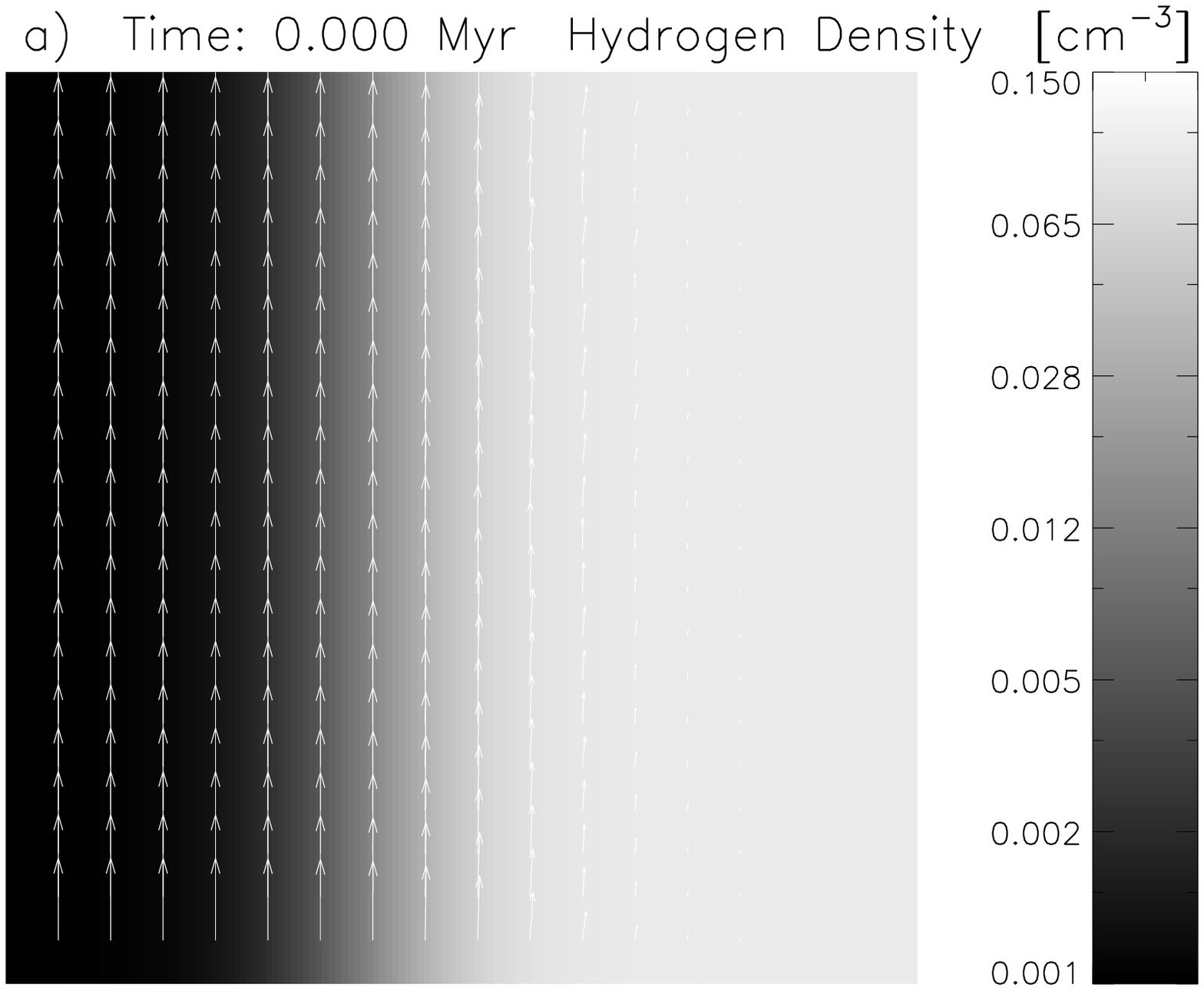}{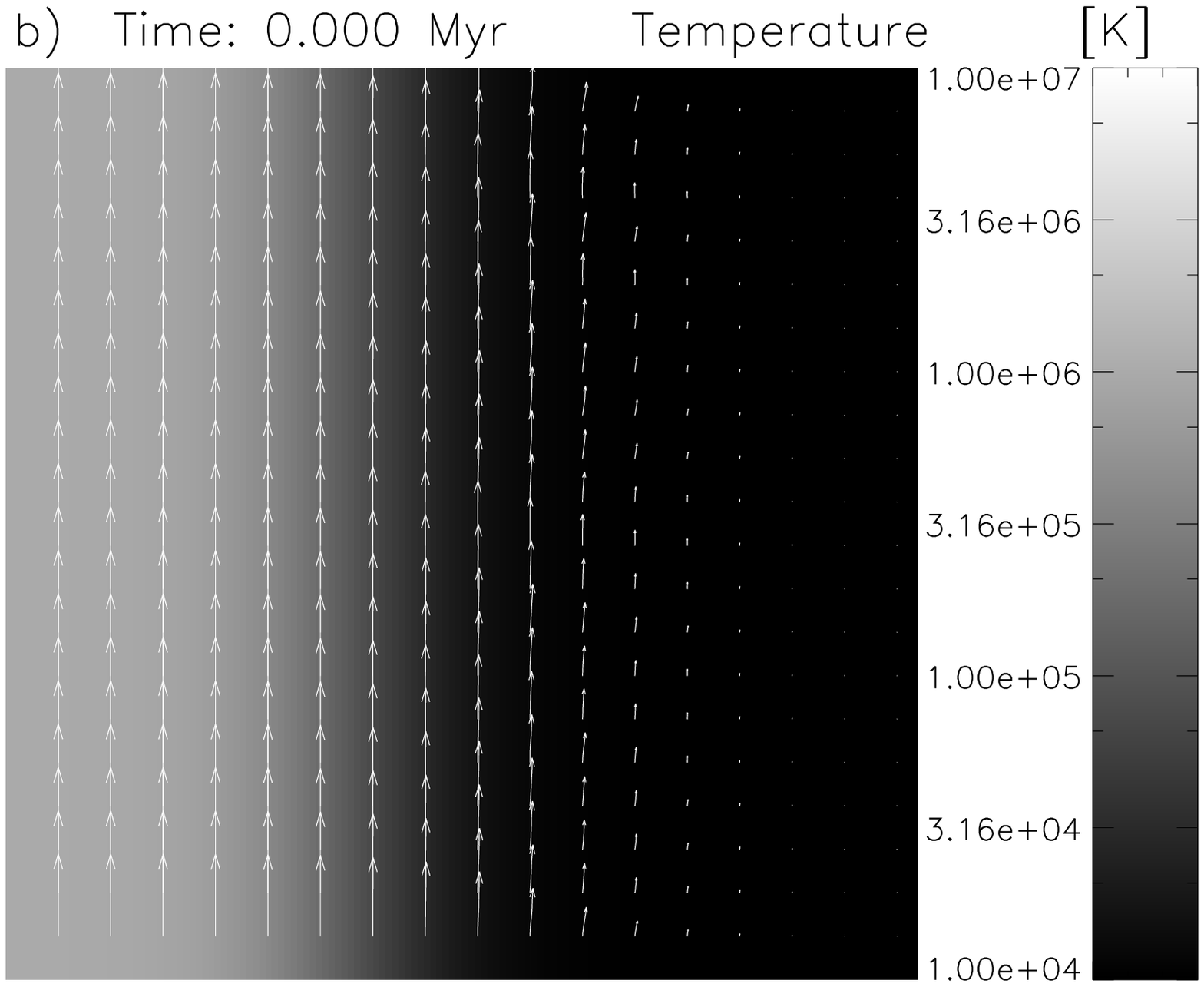}\\
\plottwo{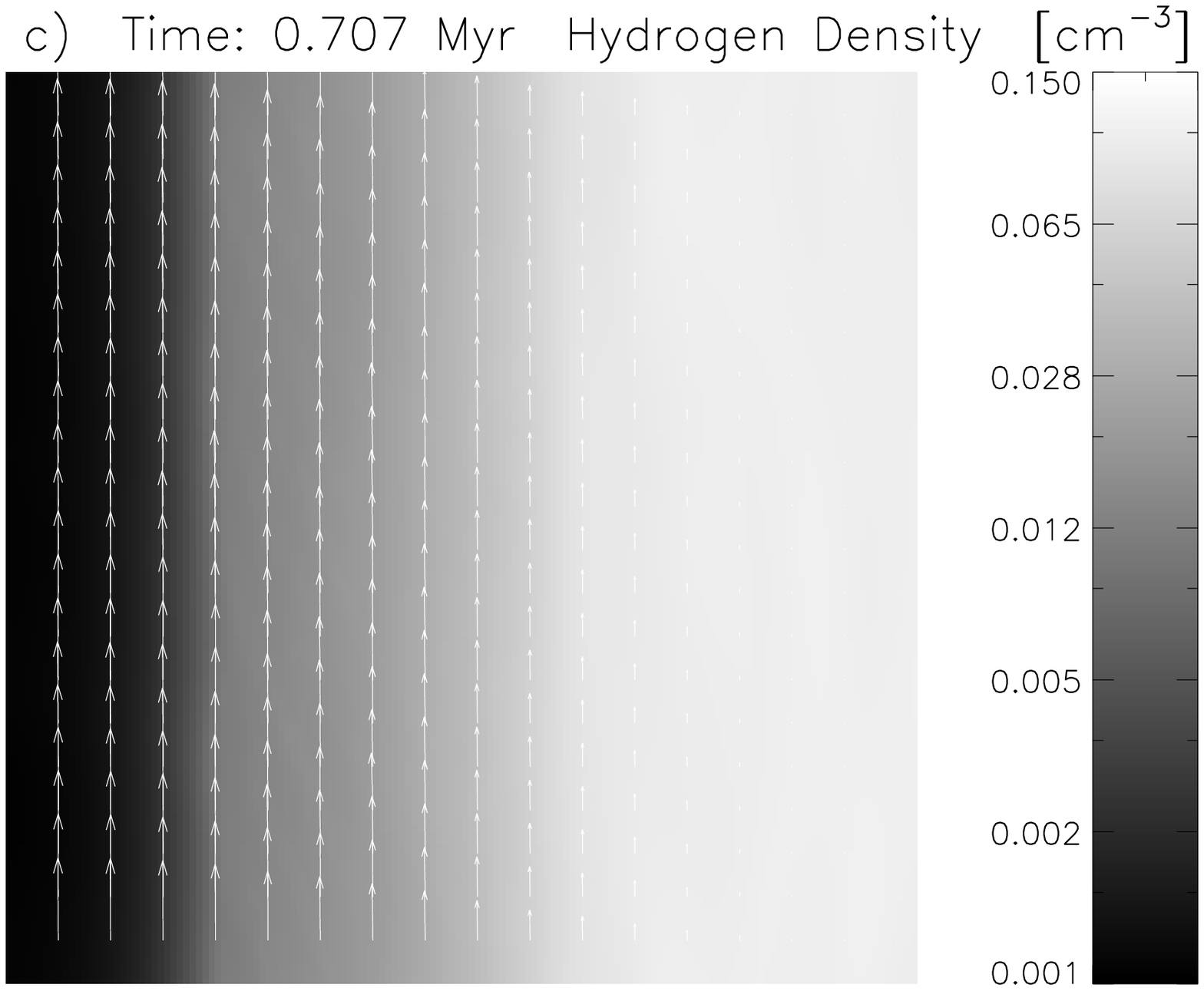}{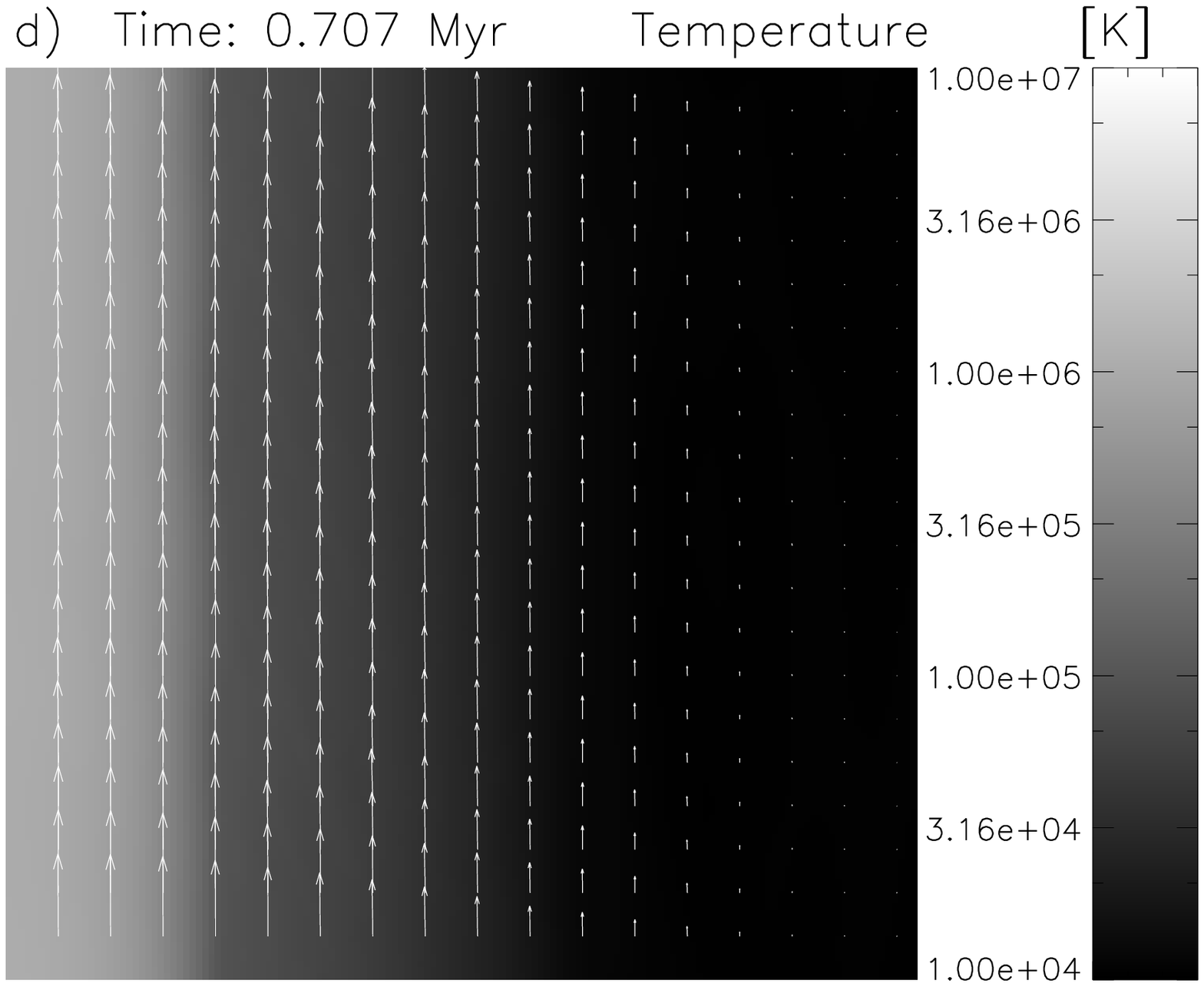}\\
\plottwo{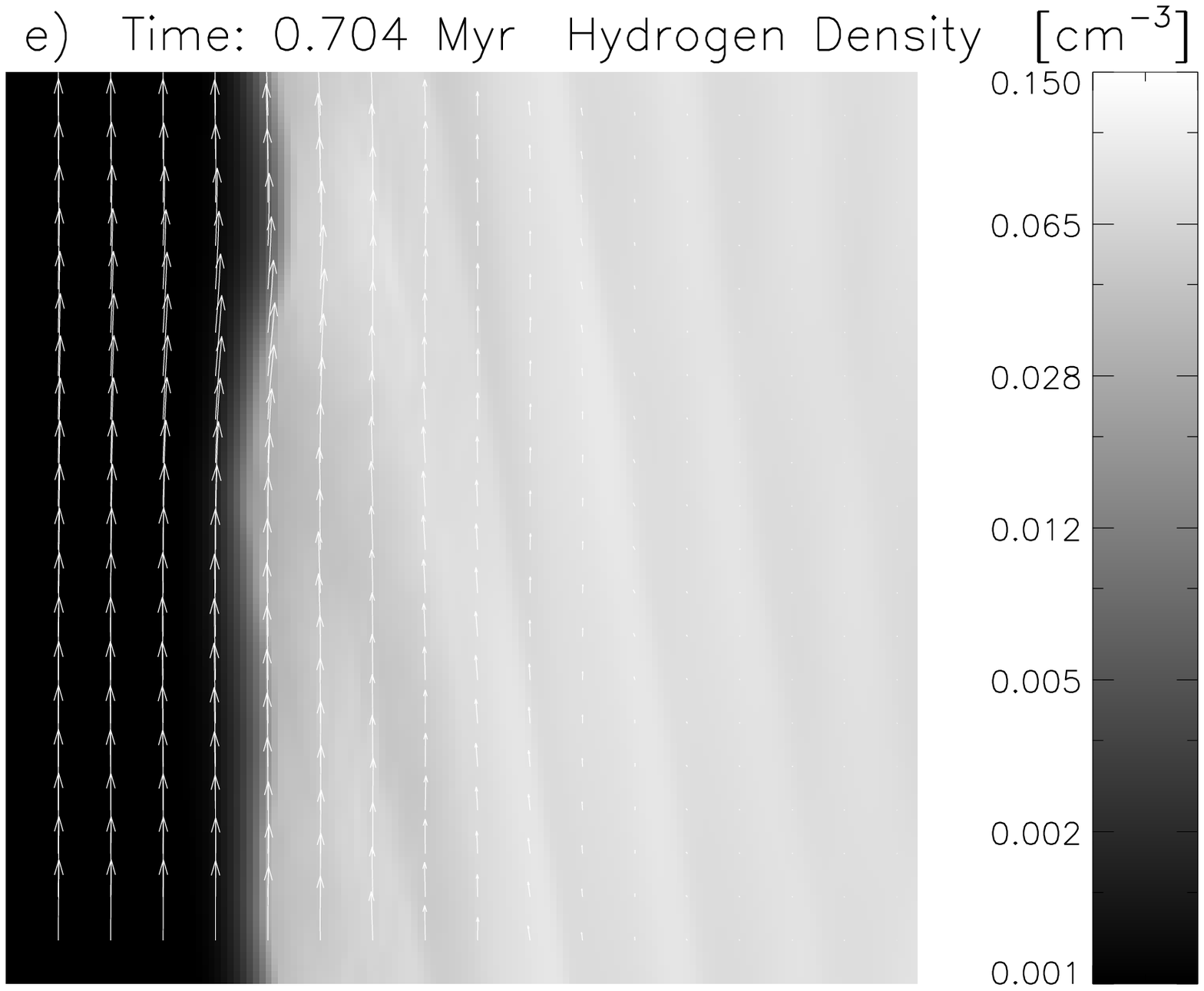}{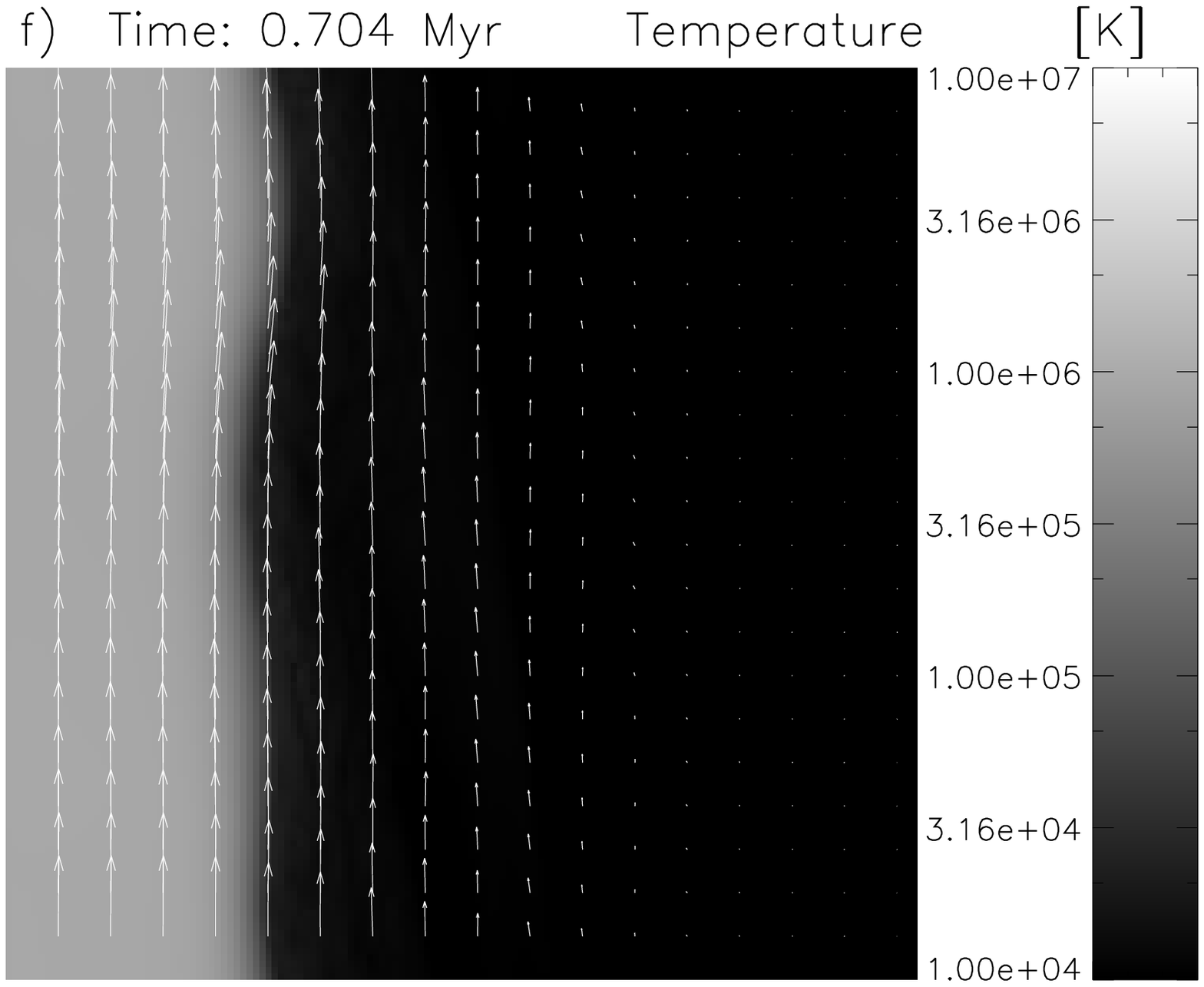}
\caption{Selected snapshots of cuts in the $XZ$ plane from our
  simulations, showing density and temperature maps, along with the
  velocity field. The largest arrow in the velocity
  representation correspond to a magnitude of $\sim
  200~\rm{km~s^{-1}}$ The upper two figures (panels {\it a, b})
  correspond show the initial conditions for $v_t=200~\rm{km~s^{-1}}$,
  and $T_{hot}=10^6~\rm{K}$. In the middle (panels {\it c, d}) we show
  the evolution after $t\sim 0.7~\rm{Myr}$, without including
  cooling. At the bottom (panels {\it e, f}) we show the result after
  approximately the same time, but with the cooling included. We can
  see how the thickness of the mixing layer grows from the initial
  conditions where no cooling is present. At the same time, we can see
  the dramatic effect of cooling showing a rather sharp transition
  zone, and also how the largest wavelength modes of the K-H
  instability start to develop.
\label{fig:snapshots1}}
\end{figure*}

Figures \ref{fig:t_ev_b0}{\it b}, and
\ref{fig:t_ev_b1}{\it b} show how the mixing layer develops.
When we do not include cooling, the
thickness of the boundary layer increases monotonically.
The growth rate is faster when the shear velocity is larger,  as
expected for the K-H instability. For the cases in which
cooling is included,  we see a very different evolution, with the
boundary layer shrinking with time at the beginning, and after some
time starting to grow again. The
smallest perturbations in the K-H instability have the fastest
growth rate, but do not have the required energy to pull (enough)
material into the  mixing interface.
Since the cooling is very effective, the transition layer is rather
sharp. At this point, we mostly have warm gas being condensed at the
interface, but mixing is not very effective yet. 
This is demonstrated in Figure \ref{fig:snapshots1}, where we show
cuts perpendicular to the mixing layer ($XZ$ plane) in two of our
simulations after $t \sim 0.7~\rm{Myr}$.
The cases with magnetic field do not show significant differences in
the formation of the mixing layer when compared with the unmagnetized
runs at early times ($t \lesssim 0.7$ Myr).

\begin{figure}
\plotone{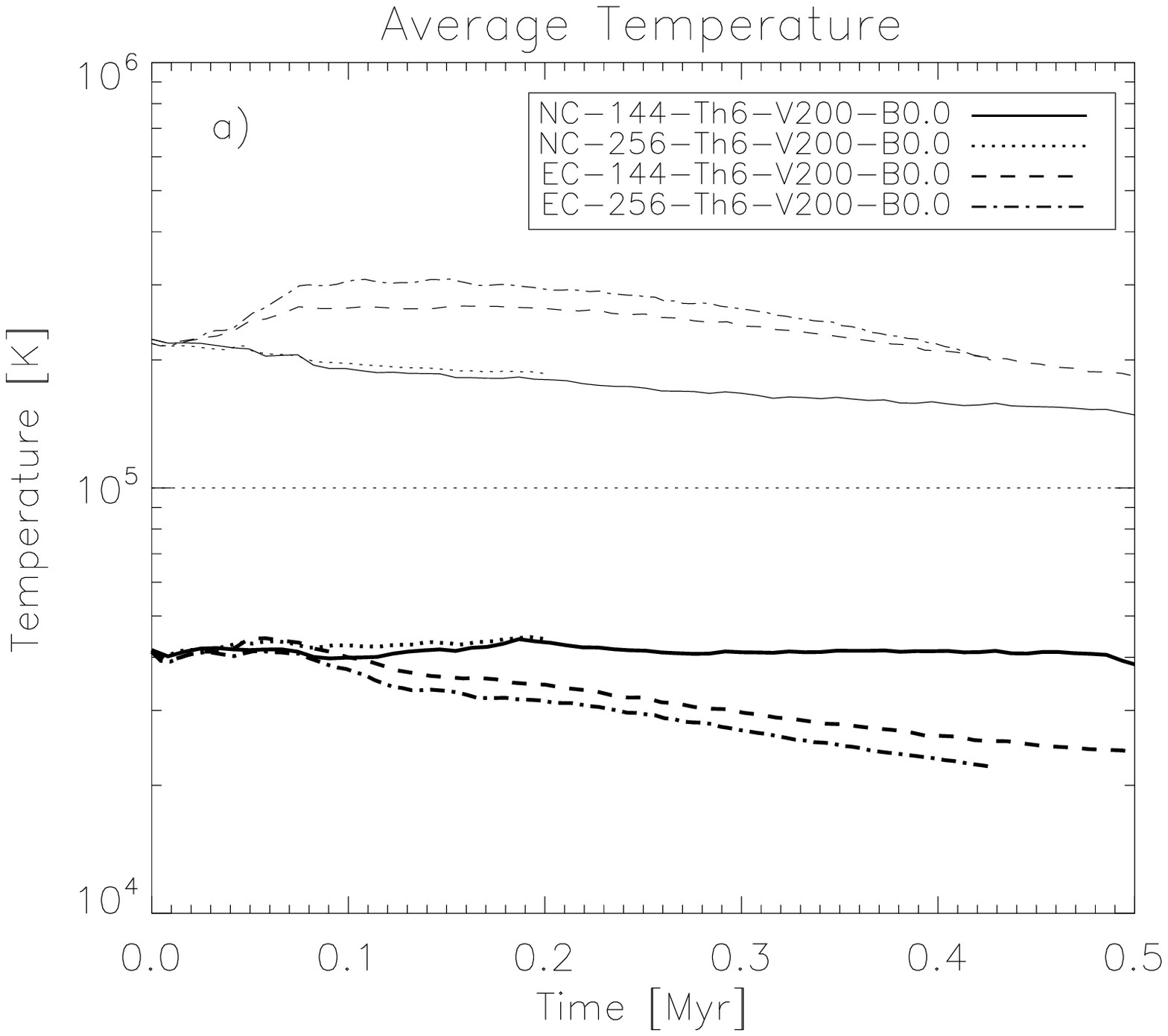}
\plotone{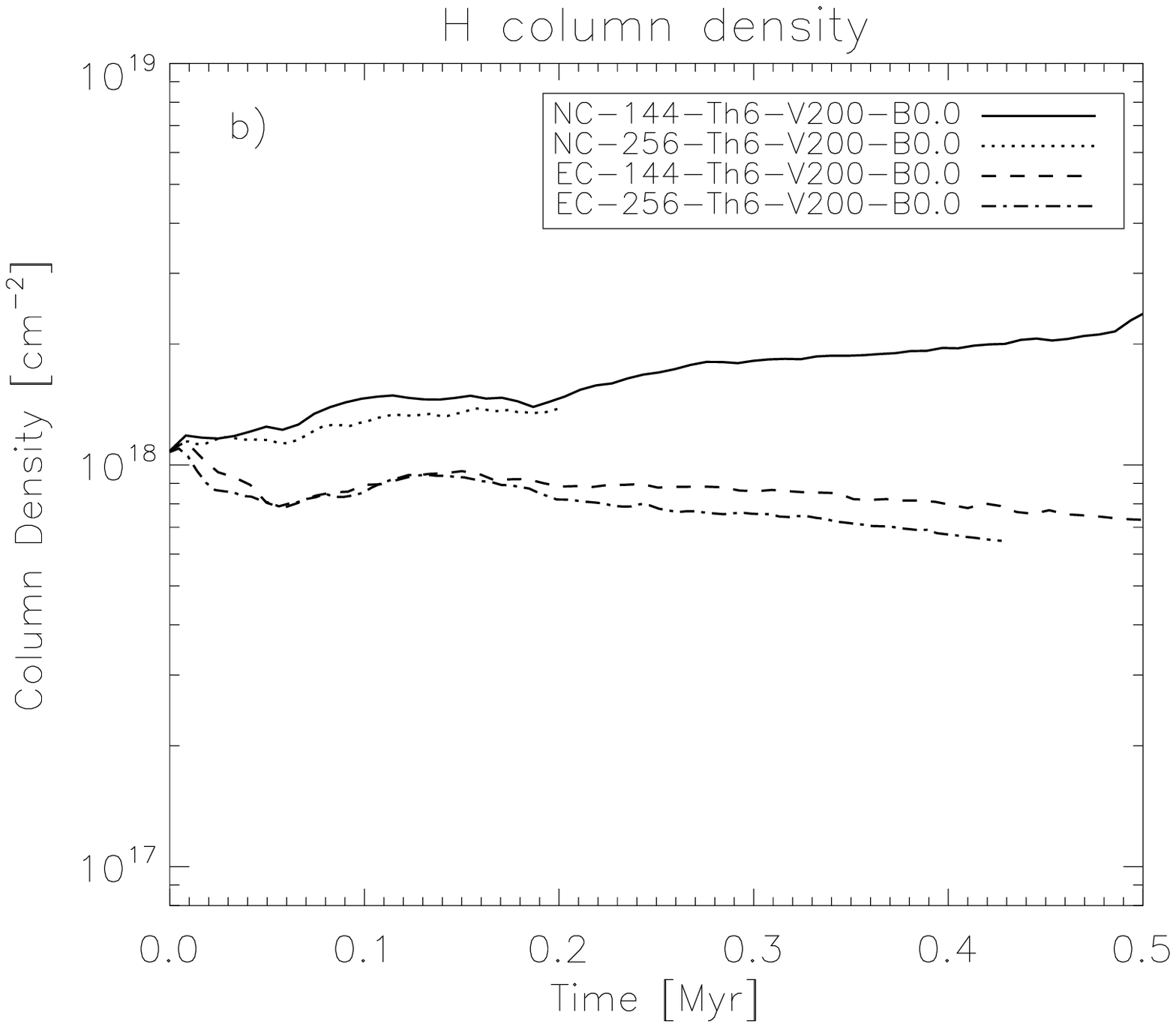}
\caption{Test of resolution convergence for a two set of
  simulations, one without cooling, and one with equilibrium
  cooling, both correspond to $T_{hot}=10^6~\rm{K}$ and
  $v_t=200~\rm{KM~s^{-1}}$. In panel {\it a}) we show the density
  weighted average temperature ({\it thick} lines), and the volume
average temperature ({\it thin} lines). Panel {\it b}), shows the
average H column density in the mixing layer.
\label{fig:hi_res}}
\end{figure}

In order to test the sensitivity of our simulations to resolution we
ran a couple of cases in a $256^3$ grid. The results are shown in
Figure \ref{fig:hi_res}.

We found marginal dependence on  numerical resolution for the
formation of the mixing layers. We will see however, that the mixing
at later times is utterly dominated by the largest scales, when the
modes that correspond to the longest wavelengths excited by the
instability become apparent.

\subsubsection{Evolution of the mixing layers at later times}

\begin{figure}
\plotone{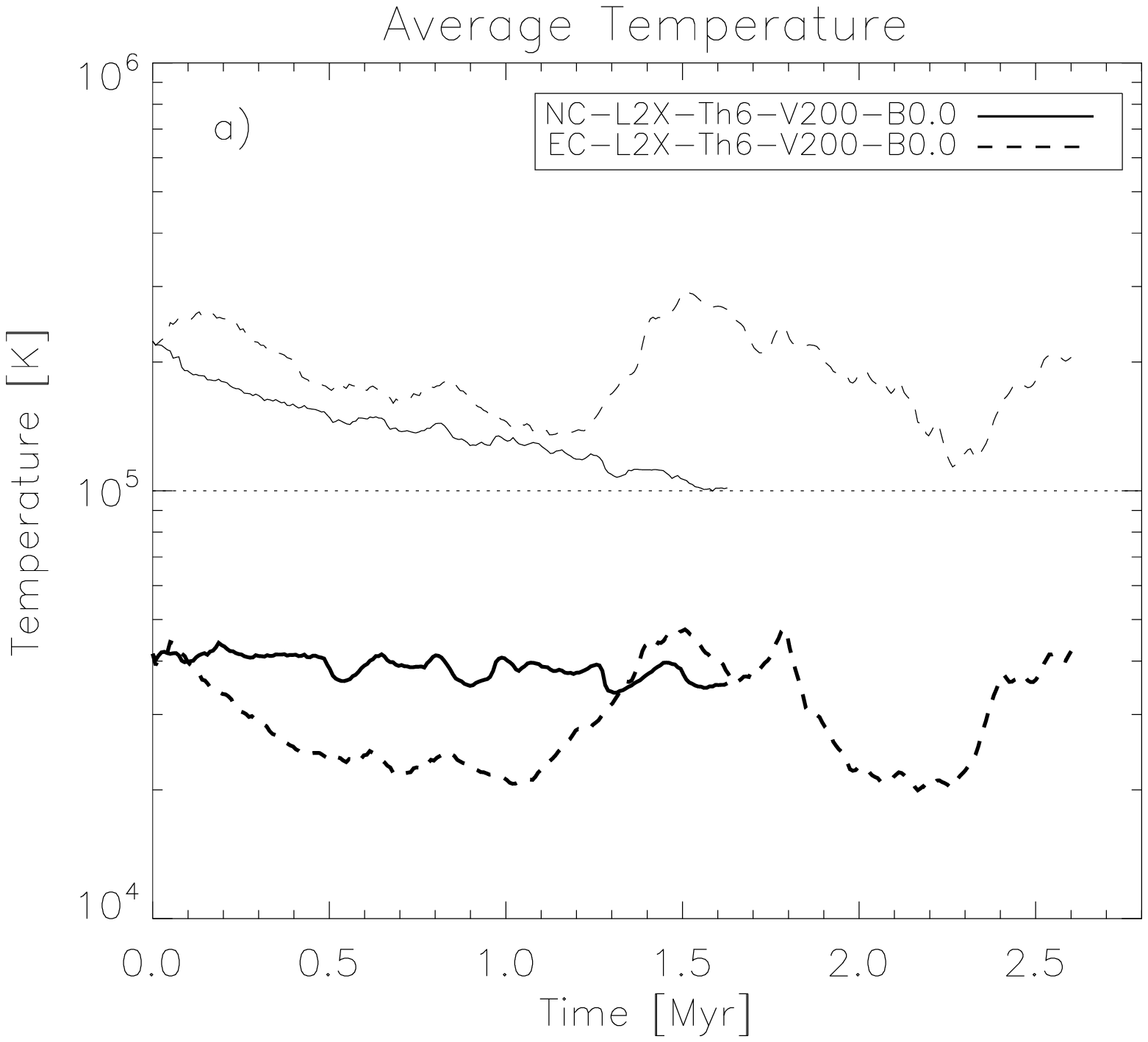}
\plotone{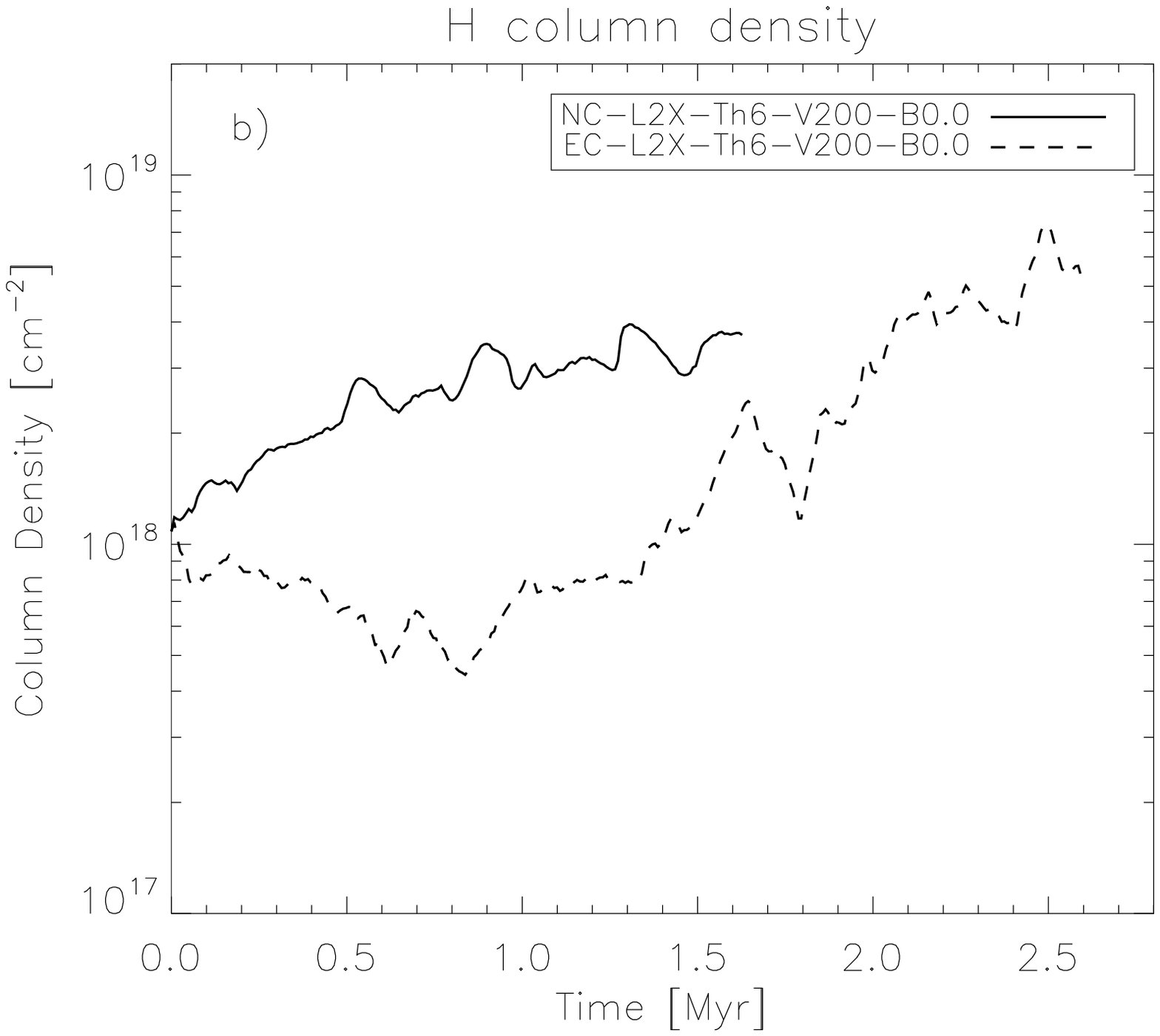}
\caption{Further evolution of the runs corresponding to
  $T_{hot}=10^6~\rm{K}$, and $v_{z}=200~\rm{m~s^{-1}}$. The top panel
 (panel {\it a}) with the average temperature, ({\it thin} lines for
  volume average, and {\it thick} lines for the density weighted.) At
  the bottom (panel {\it b}) we see that for the case with equilibrium
  cooling, after the initial stage, the the mixing layer starts to
  grow.
\label{fig:2LX}}
\end{figure}

Up until now, we have seen the formation of turbulent mixing layers
by means of a K-H instability. However, we have not seen evidence of
reaching a steady state, which is the original idea of the whole
process.
Since the largest modes in the K-H instability grow rather slowly, it
is not practical to follow the evolution of all our models until
steady state is achieved. Moreover, in a qualitative way, we have
observed similar behavior for the runs that started with a larger
temperature ($T_{hot}=10^7~\rm{K}$ ), but with the additional
difficulty that we require finer time stepping for such cases, making
it very difficult to cover a large span in time. 
To study the long-term evolution of the layer, we followed the
the fastest evolving model. That is, the unmagnetized,
$T_{hot}=10^6~\rm{K}$, and $v_t=200~\rm{km~s^{-1}}$ model.
A related problem in our $144^3$ resolution runs was that by the time
the mixing layer was reaching its asymptotic state (after $\sim 1.5
~\rm{Myr}$), it was getting very close to the boundary of the
computational box. To overcome this problem we extended the dimensions
of the box to $256$ cells in the $x$ direction, and we placed the
division between hot and warm media off-center.
The new initial conditions (``LX-Th6-V200-B0'' in Table
\ref{tb:param}) have only $1/4$th of the volume filled with warm gas,
and the remainder with hot gas. This is because we have much more heat
capacity in the warm gas (due to the density contrast), and because our
simulations did not have enough hot gas to provide a steady state.
Figure \ref{fig:2LX} illustrates the evolution of this extended
model, which shows how the mixing layer broadens after the first signs
of large modes of the instability appear (see Figures
\ref{fig:snapshots1}{\it e} and \ref{fig:snapshots1}{\it f}).

\begin{figure*}
\plotone{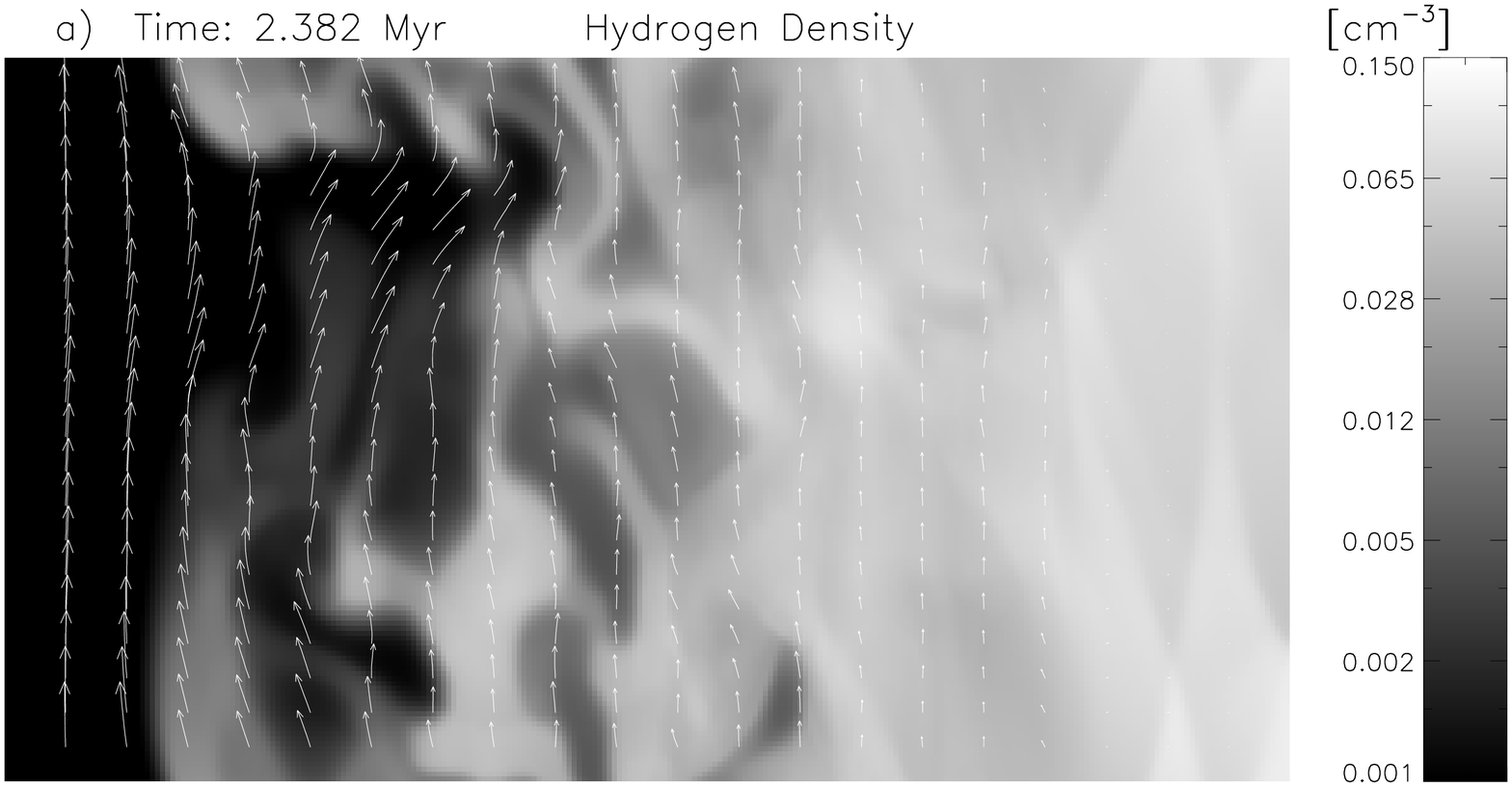}\\
\plotone{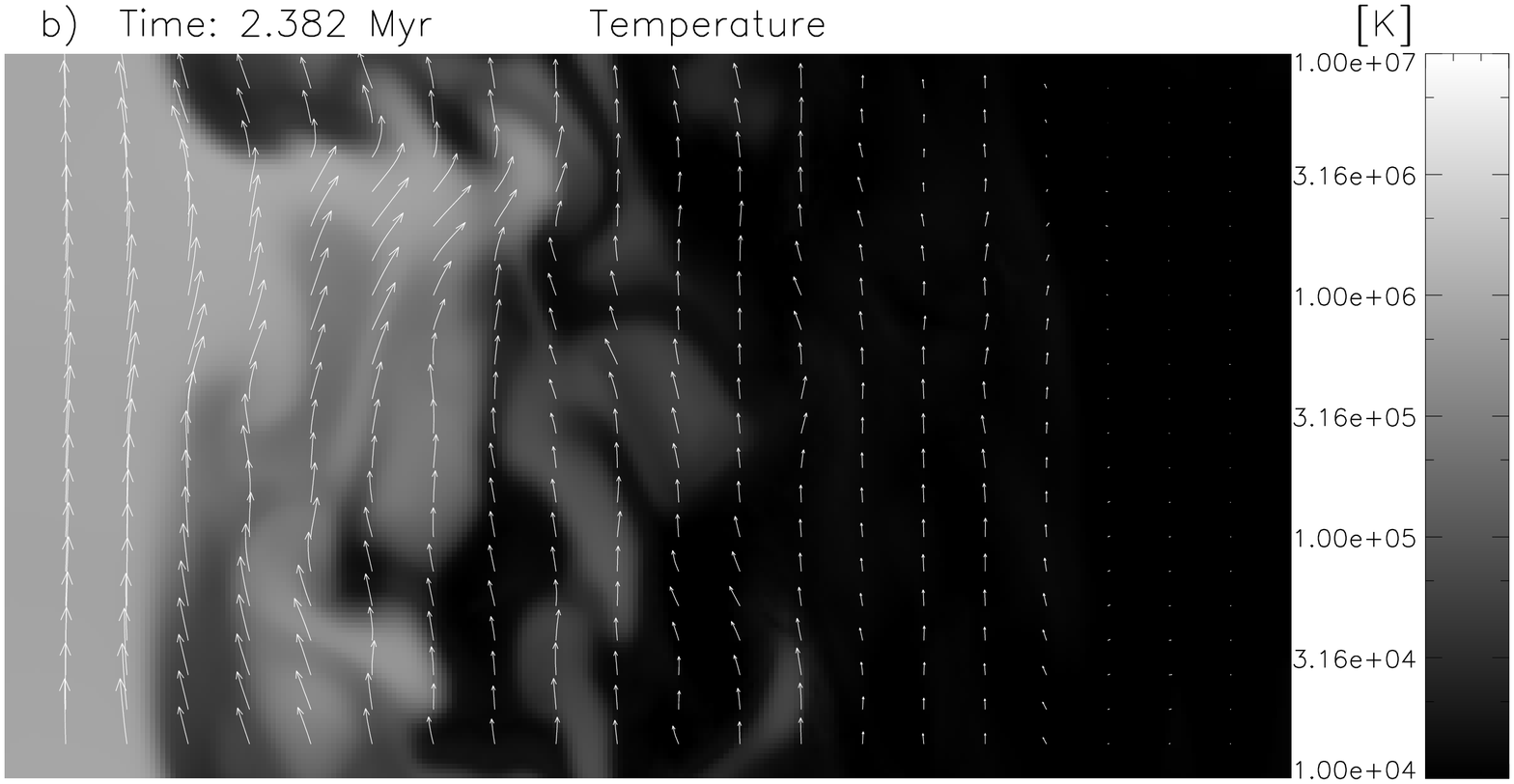}
\caption{ Snapshots of cuts in the $XZ$ plane for ``NC-L2X-Th6-V200-B0.0''
  after $t\sim 2.3~\rm{Myr}$. Notice how the K-H instability has
  indeed produced turbulence at the boundary of the two media,
  increasing the thickness of the zone with intermediate temperatures
  ($T \sim 10^5~\rm{K}$).
\label{fig:snapshots2}}
\end{figure*}

We also show snapshots of cuts perpendicular to the layer, after
$2.3~\rm{Myr}$ in Figure \ref{fig:snapshots2}. This figure highlights 
how a fully operational K-H instability effectively mixes the two media
and provides a much larger interface zone, compared with the times
when only small scale modes are present.
Even after $\sim 2.5~\rm{Myr}$, in the longer box ($256$ cells in the
$x$ direction), the simulations have not reached a steady
state. Nonetheless, our results provide important insights on how
these mixing layers will develop and change over time.

\subsection{Observational Diagnostics for Mixing Layers}

\begin{figure}
\epsscale{1.25}
\plotone{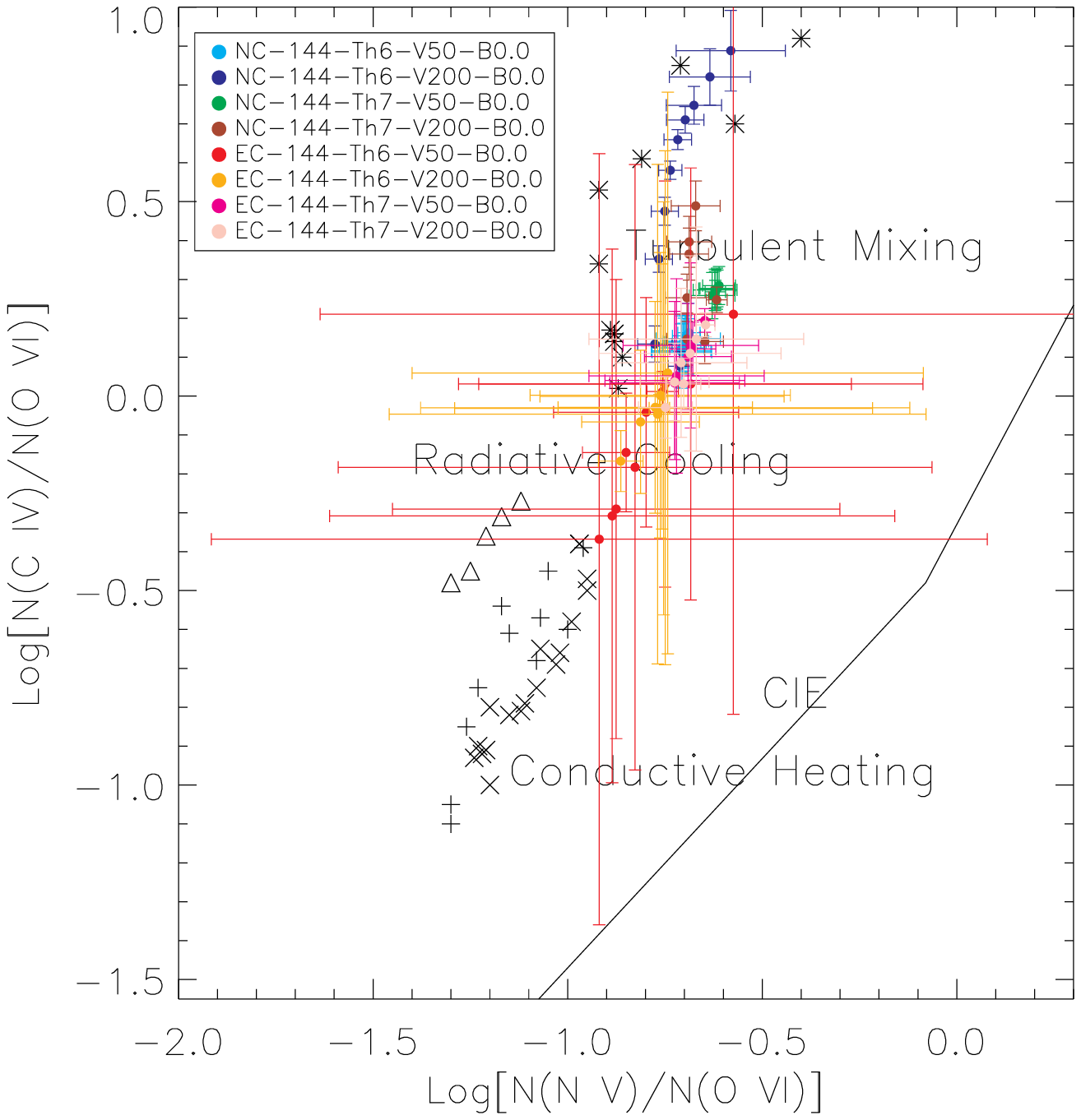}
\caption{Comparison of line ratios for various models in the
  literature, and our calculations (modified
  from \citet{IS04a,IS04b}. The {\it triangles} correspond to radiative
  cooling from a Galactic fountain model \citep{SB93,BS93}. The
  ``x's'' are conductive heating and evaporation models of planar 
  clouds \citep*{BH87,BBF90}. Cooling of SNR shells are represented by
  ``+'s'' \citep{SC93,S98}. The ion ratios in collisional ionization
  equilibrium from \citet{SD93} are in {\it solid} line (this
  corresponds to a range in temperatures, higher at the bottom center
  and lower at the upper right). SSB83 result for turbulent mixing
  layers are in {\it asterisks}. Our calculations are represented with
  {\it circles},  the color coding can be found in the legend of the
  plot. These last correspond to the average of LOSs normal to the
  boundary, for a given point in time each. The error bars were
  obtained considering the variability with the different LOSs.
\label{fig:ratios}}
\end{figure}

\begin{figure}
\epsscale{1.25}
\plotone{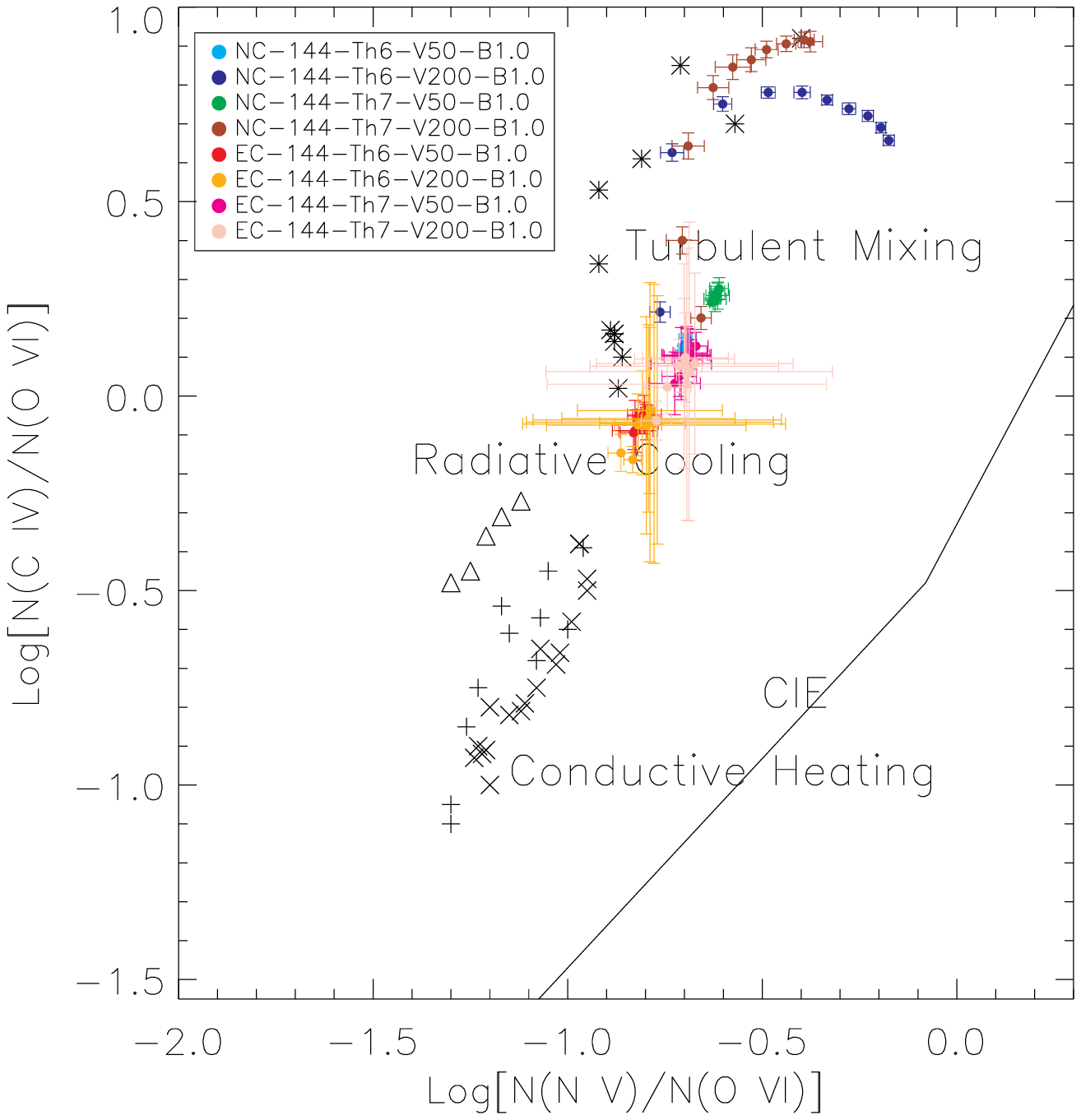}
\caption{ Same as Fig. \ref{fig:ratios}, with our simulations that
  include magnetic field ($\beta \sim 1$).
\label{fig:ratios_b1}}
\end{figure}

Notably, the time evolution we observe for these mixing layers are
associated  with a significant evolution of spectral line
diagnostics, most importantly the  column densities and 
ratios of highly ionized species. These measurements depend not only on
the metallicity, density contrast, and relative velocities of the
mixing zones, but also on the evolutionary state of the layer.  As a
first step towards examining the evolution, we have calculated the
column densities of \ion{C}{4}, \ion{N}{5}, and \ion{O}{6} ions
through our simulated layers as a function of time.  Using the
temperature and density profile along  synthetic lines-of-sight (LOSs)
through the data cube, chosen perpendicularly to the boundary layer
(in the direction of the $x$ axis). These column densities were
computed under the assumption of solar metallicity, and collisional
ionization equilibrium \citep{BBC01}.
Figures \ref{fig:ratios} and \ref{fig:ratios_b1} show the range of
values of the \ion{C}{4}/\ion{O}{6}, and \ion{N}{5}/\ion{O}{6} ratios
along with  other models of high ion production in the literature
(compiled by \citealt{IS04a,IS04b}). The central points are the mean
\ion{C}{4}/\ion{O}{6}, and \ion{N}{5}/\ion{O}{6} values averaged over
all the possible LOSs, at a given point in time. The error bars were
obtained at each time, with the dispersion of column densities for the
various LOSs. The time domain was splitted evenly into $10$ bins from
$t=0$ to the longest time available for each run (the final times can
be checked in Figures \ref{fig:t_ev_b0}, \ref{fig:t_ev_b1}, and
\ref{fig:2LX}).

The line ratios and column densities from our simulations should not
be interpreted too literally, but rather as a guide to obtain insight
into the evolution of mixing layers.
In particular, the assumption of collisional ion fraction equilibrium
limits the accuracy of line ratios. 
As pointed out by SSB93, mixing layers can be quite far
from equilibrium.
The overall dynamics and time evolution of mixing layers is not likely
to change dramatically  by non-equilibrium effects. However,
observational diagnostics such as line ratios can be affected
significantly.
A study of non-equilibrium cooling models with more emphasis on the
observational implications is necessary and we plan to provide it in
a following paper.

It is interesting that a clear difference is present between the runs
that include cooling and those that do not.
However, note that except for the ``LX-Th6-V200'' runs, the mixing
layer is still in a very early stage  of formation.
It is remarkable that the models without cooling show similarity to
the results of SSB93. Certainly their models did not account for the
dynamical effect of the cooling: their model provides the
intermediate temperatures, but the mixing is merely hydrodynamical.
Condensation or evaporating flows were not considered. At the same
time, their model also assumed fully developed turbulence, and our set
of runs have not achieved the corresponding stationary state.

With new models, observations may be able to provide insights about the
time-dependence of the mixing.
Another thing to notice is the larger scatter for the cases that have
been run for longer times. 
This scatter can be explained noting that for longer timescales we
have a more dynamical (turbulent) picture, with many structures moving
in and out of a particular LOS. Actually, the values of the
\ion{C}{4}/\ion{O}{6}, \ion{N}{5}/\ion{O}{6} ratios (and the rather
large scatter), are somewhat consistent with the observations
presented by \citet{IS04b}.
Is is also evident that in the magnetized cases the K-H instability
(and therefore the development of turbulence at the boundary) is
delayed. This can be seen from the smaller scatter of  points in
Figure \ref{fig:ratios_b1} compared to  those in \ref{fig:ratios}.
However, magnetic reconnection  should not allow the magnetic field to
form knots and prevent turbulent mixing motions \citep[see][]{LV99}.

It remains difficult to reconcile the models with high ion column
densities. It has been pointed out (e.g. SSB93;
\citealt{S03,IS04a,IS04b}), that to explain the typical column
densities observed,the LOS must pass through several mixing layers
(sometimes as many as a hundred). 
Indeed, several interfaces are likely to be blended for long
lines-of-sight (as in the case of observations in the halo of the
galaxy). However, in many cases fairly smooth and symmetrical line
profiles, with little centroid dispersions are observed.
This evidence indicates that merely summing over a large number of
interfaces may not explain the observed column densities.
In Table \ref{tb:columns} we present the column densities of an
arbitrary artificial LOS after $0.7$ Myr for all the
models with $T_{hot}=10^6~\rm{K}$. For comparison we include the
values that correspond to the initial conditions.
\begin{deluxetable*}{lcccccccc}
\tabletypesize{\scriptsize} 
\tablecaption{Ion Column Densities$^*$ at early times.
\label{tb:columns}}
\tablecolumns{12}
\tablewidth{0pt}
\tablehead{
       \colhead{Model} & \multicolumn{2}{c}{\ion{C}{4}} & \colhead{} &
                         \multicolumn{2}{c}{\ion{N}{5}} & \colhead{} &
                         \multicolumn{2}{c}{\ion{O}{6}} \\
\cline{2-3} \cline{5-6} \cline{8-9} \\
\colhead{} & 
\colhead{$t=0$ Myr.}& \colhead{$t=0.7$ Myr.} &&
\colhead{$t=0$ Myr.}& \colhead{$t=0.7$ Myr.} &&
\colhead{$t=0$ Myr.}& \colhead{$t=0.7$ Myr.}
}
\startdata
NC-144-Th6-V50-B0.0 &  $12.03$ & $12.06$  &&
                       $11.19$ & $11.20$  &&
		       $11.89$ & $11.90$  \\
EC-144-Th6-V50-B0.0 &  $12.03$ & $11.53$  &&
                       $11.19$ & $10.74$  &&
                       $11.89$ & $11.45$  \\
NC-144-Th6-V200-B0.0 & $12.03$ & $12.34$  &&
                       $11.19$ & $11.05$  &&
                       $11.89$ & $11.79$  \\
EC-144-Th6-V200-B0.0 & $12.03$ & $11.26$  &&
                       $11.19$ & $10.55$  &&
                       $11.89$ & $11.33$  \\
NC-L2X-Th6-V200-B0.0 & $12.03$ & $12.36$  &&
                       $11.20$ & $11.06$  &&
                       $11.93$ & $11.83$  \\
EC-L2X-Th6-V200-B0.0 & $12.03$ & $11.38$  &&
                       $11.20$ & $10.67$  &&
                       $11.93$ & $11.57$  \\
NC-144-Th6-V50-B1.0  & $12.29$ & $12.34$  &&
                       $11.47$ & $11.51$  &&
                       $12.18$ & $12.22$  \\
EC-144-Th6-V50-B1.0  & $12.29$ & $11.61$  &&
                       $11.47$ & $10.85$  &&
                       $12.18$ & $11.71$  \\
NC-144-Th6-V200-B1.0 & $12.29$ & $12.84$  &&
                       $11.47$ & $11.65$  &&
                       $12.18$ & $12.06$  \\
EC-144-Th6-V200-B1.0 & $12.29$ & $11.50$  &&
                       $11.47$ & $10.76$  &&
                       $12.18$ & $11.58$ 
\enddata
\tablenotetext{*}{Given as log of the ion column density in
  cm$^{-2}$.}
\end{deluxetable*}

\begin{figure}
\plotone{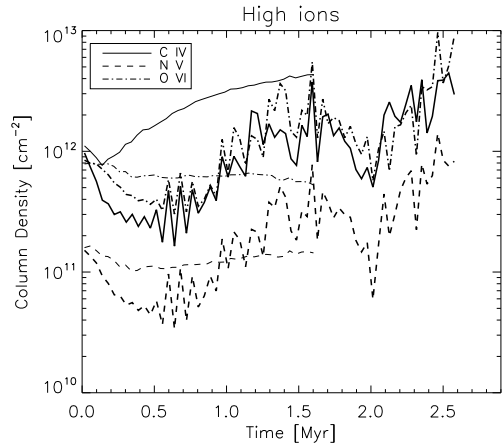}
\caption{Time evolution of the column density of different high
  ions for a line of sight perpendicular to the interface. The
  initial conditions correspond to $T_{hot}=10^6~\rm{K}$ and
  $v_t=200~\rm{km~s^{-1}}$.{\it Thin} lines correspond to the
  case without cooling, {\it thick} lines to equilibrium cooling.
\label{fig:column}}
\end{figure}

In Figure \ref{fig:column} we show the mean column densities of
\ion{C}{4}, \ion{N}{5}, and \ion{O}{6} (for synthetic LOSs normal to
the interface), as a function of time for the more evolved runs.
Average values of column densities in the halo of the Galaxy observed
with {\it FUSE} are \citep{S03,IS04b}: $\log N [$\ion{C}{4}$]\sim
14.3$, $\log N [$\ion{N}{5}$]\sim 13.7$, and $\log N
[$\ion{O}{6}$]\sim 14.5$.
A survey with the {\it Copernicus} satellite in the Galactic plane by
\citet{J78a,J78b}, suggested that individual \ion{O}{6} components
have column densities closer to $\log N [$\ion{O}{6}$]\sim 13$, which
is similar to observations with {\it FUSE} of very nearby stars
\citep{OJSBC05}.
The column densities obtained in the model that ran for the
longest time (at $t\sim 2.5$ Myr), are slightly larger than those
predicted by SSB93, but would still require too many interfaces to
explain observations. However, as we have discused earlier, the
simulations were stopped before fully reaching a stationary state, and
the thickness of intermediate temperature zone was still increasing
with time (see for instance Figure \ref{fig:2LX}{\it b}).
The results presented here suggest that not only the number of mixing
layers, but the time available for turbulence to develop, are factors
to consider for the proper interpretation of observations.

\section{Summary}

Turbulent mixing layers at the interfaces of the hot ($T\sim
10^{6-7}~\rm{K}$) and warm ($T\sim 10^4~\rm{K}$) media in the ISM
can be produced by high shear, via a Kelvin-Helmholtz instability.
We use a MHD code with radiative cooling, to model the formation and
evolution of turbulent mixing layers through this instability.

The introduction of cooling in to the dynamical evolution was found to
produce dramatic differences compared with the same models ran without
cooling. For instance, the deposition of momenta from hot gas
condensing onto the mixing layer modifies the growth-rate of the
K-H instability, making it develop faster than the same case run
without cooling. 

The low density of the hot medium caused the growth rate of
the large scale modes of the instability to be slow. At early
stages of the formation of the mixing layer ($t~\lesssim 1~\rm{Myr}$)
the rapid cooling of material at intermediate temperatures ($T\sim
10^5~\rm{K}$) makes for a sharp transition between the hot and warm
media.

At later times, the instability excites large scale fluctuations that
are powerful enough to provide more efficient mixing, and the
transition zone broadens.
We see evidence that for typical ISM conditions ($T_{hot}\sim
10^6~\rm{K}$ and $T_{warm}\sim 10^4~\rm{K}$, with a shear velocity of
  $200~\rm{km~s^{-1}}$), the mixing layer is still growing after $2.5
  \rm{Myr}$.

We included a dynamically important magnetic field ($\sim
2~\mu\rm{G}$) in the warm phase for some runs, and its effect was
found to be minimal at earlier times $\lesssim 0.7$ Myr. At later
times the magnetic field inhibits/delays the development of the K-H
instability, and thus the turbulent mixing.

Assuming solar metallicities, and collisional ionization equilibrium
fractions, we estimated column densities and line ratios of highly
ionized species (\ion{C}{4}, \ion{N}{5}, and \ion{O}{6}) from our
simulations.
We compared our results with previous models (SSB93) of turbulent
mixing layers and found similar column densities. The
\ion{C}{4}/\ion{O}{6}, and \ion{N}{5}/\ion{O}{6} mean line ratios are
slightly different, with a lower \ion{C}{4}/\ion{O}{6}. We also found
a significant scatter between different lines-of-sight, in particular
for the models evolved to longer times.

We were not able to fully reach a stationary state in our simulations,
however our results suggest that given more time ($\gtrsim 2.5
~\rm{Myr}$ for typical conditions), the column density of high ions
should be significantly larger than previous models, partially
alleviating the problem of number of layers required to explain
observations.

\acknowledgements{This work was partially supported by the National
  Computational Science Alliance under grant AST050010N and utilized
  the Xeon Linux Cluster. AE acknowledges financial support from
  CONACYT (Mexico). AL acknowledges Space Telescope Theory Grant
  HST-AR-09939.01, and the NSF Center for Magnetic Self Organization
  in Laboratory and Astrophysical Plasmas. SNL participation was
  possible thanks to the Research Experience for Undergraduates
  program at the University of Wisconsin-Madison.}

%

\end{document}